\documentclass[superscriptaddress,twocolumn,showpacs,prb]{revtex4}
\newcommand{\figurewidth}{\columnwidth}
\newcommand{\av}{_{\rm av}}
\newcommand{\z}{_{\rm z}}
\newcommand{\smfrac}[2]{\mbox{\small $#1 \over #2$}}
\usepackage{amsmath,graphicx,tabularx}
\usepackage{graphicx,amsmath}

\begin{document}

\title{
The de Almeida-Thouless line in vector spin glasses}

\author{Auditya Sharma}
\affiliation{Department of Physics, University of California,
Santa Cruz, California 95064}
\author{A.~P.~Young}
\email{peter@physics.ucsc.edu}
\affiliation{Department of Physics, University of California,
Santa Cruz, California 95064}

\date{\today}

\begin{abstract}
We consider the infinite-range spin glass in which the spins have $m>1$
components (a vector spin glass). Applying a magnetic field which is random
in direction, there is an Almeida Thouless (AT) line below which the "replica
symmetric" solution is unstable, just as for the Ising ($m=1$) case.  We
calculate the location of this AT line for Gaussian random fields for
arbitrary $m$, and verify our
results by numerical simulations for $m = 3$.
\end{abstract}
\pacs{75.50.Lk, 75.40.Mg, 05.50.+q}
\maketitle

\section{Introduction}
\label{sec:intro}

The infinite-range Ising spin glass, first proposed by Sherrington and
Kirkpatrick~\cite{sherrington:75}, has been extensively studied. It was found
by de Almeida and Thouless\cite{almeidaAT:78} (hereafter referred to as AT)
that 
the simple ``replica symmetric'' (RS) ansatz for the spin glass state becomes
unstable below a line in the magnetic field-temperature plane, known as the
AT line. While the Ising spin has $m=1$ components,
the $m$-component vector spin glass for $m>1$ has received less
attention. de Almeida et
al.\cite{almeidaAT:78b} (hereafter referred to as AJKT)
found an instability in zero field, but did not
consider the effects of a magnetic field. The effects of a \textit{uniform}
field on a
vector spin glass were first studied by Gabay and Toulouse\cite{gabay:81}.
They found a line of transitions (the GT line),
which is of a different nature from
the AT line. In a uniform field, a distinction has to be made between spin
components longitudinal and transverse to the field, and the GT line is the
spin glass ordering of the transverse components, and these are effectively in
\textit{zero field}\cite{moore:82,cragg:82}.
The AT line is
different from the Gabay-Toulouse\cite{gabay:81} (GT) line, since it is a
transition to a phase with replica
symmetry breaking but with \textit{no change in
spin symmetry}. The existence of the AT line is perhaps the most striking
prediction of the mean field theory of spin glasses. The GT line occurs at a
higher temperature than the putative AT line, which becomes simply a
crossover\cite{moore:82,cragg:82}.

The main point of the present work is to argue
that one should consider not a uniform
field but a field which is random in \textit{direction} (it will also be
convenient to make it random in magnitude though this is not essential) and
that, in this case,
there \textit{is} an AT line also for vector spin glasses. We
will
determine the location of this line for
an arbitrary number of spin components.

The Hamiltonian is given by
\begin{equation}
\mathcal{H} = -\sum_{\langle i, j \rangle} J_{ij} \mathbf{S}_i \cdot \mathbf{S}_j -
\sum_i \mathbf{h}_i \cdot \mathbf{S}_i \, ,
\label{Ham}
\end{equation}
where the $S_i^\mu,\ (\mu = 1, \cdots, m)$ are $m$-component spins of
length $m^{1/2}$, i.e.
\begin{equation}
\sum_{\mu = 1}^m \left(S_i^\mu\right)^2 = m, 
\label{norm}
\end{equation}
the interactions $J_{ij}$ between all distinct pairs of spins $\langle i,
j \rangle$ are independent random variables with zero mean and variance
given by
\begin{equation}
[J_{ij}^2 ]\av = {J^2 \over N - 1}\, ,
\label{Js}
\end{equation}
and the $h_i^\mu$ are independent Gaussian random fields, uncorrelated between
sites, with
zero mean and which satisfy
\begin{equation}
[ h_i^\mu h_j^\nu]\av = h_r^2\, \delta_{ij}\, \delta_{\mu\nu} \, .
\label{hs}
\end{equation}
The notation $[\cdots]\av$ indicates an average over the quenched disorder.
The normalization of the spins in Eq.~\eqref{norm} is chosen so that the zero field
transition temperature is
\begin{equation}
T_c = J
\label{Tc}
\end{equation}
for all $m$.

Consider first the Ising case ($m=1$).
The spin glass order parameter
is 
\begin{equation}
q \equiv {1 \over N} \sum_i [\langle S_i \rangle^2 ]\av ,
\end{equation}
where $\langle \cdots \rangle $ denotes a thermal average.
From linear response theory, if we make small additional random changes,
$\delta h_i$, in the random fields, uncorrelated with each other and the
original values of the fields,
the change in $\langle S_i \rangle$ is given by  
\begin{equation}
\delta \langle S_i \rangle = {1 \over T} \sum_j \chi_{ij}\, \delta h_j \, ,
\end{equation}
where the linear response function $\chi_{ij}$ is given by
\begin{equation}
\chi_{ij} = \langle S_i S_j \rangle - \langle S_i \rangle \langle S_j \rangle
\, ,
\end{equation}
and, for convenience, we have separated out the factor of $1/T$. Hence the
change in $q$ is given by
\begin{align}
\delta q &=  {1 \over T^2} {1 \over N} \sum_{i, j, k} \Bigl[ \chi_{i j}
\chi_{i k} \Bigr]\av [\delta h_j \delta h_k]\av , \\
&= {1 \over T^2}\, \chi_{SG} \, \, \delta h_r^2 ,
\end{align}
where 
\begin{align}
\chi_{SG} &= {1 \over N} \sum_{i, j} \Bigl[ \chi_{i j}^2 \Bigr]\av,  \nonumber \\
&= {1 \over N} \sum_{i, j} \Bigl[ \left(\  \langle S_i S_j \rangle - \langle
S_i \rangle \langle S_j \rangle\ \right)^2   \Bigr]\av 
\label{chisg_ising}
\end{align}
is the spin glass susceptibility. 

The corresponding results for vector spins are easily obtained.
The
change in the spin glass order parameter,
\begin{equation}
q \equiv {1 \over N} \sum_i {1 \over m} \sum_\mu [\langle S_i^\mu \rangle^2 
]\av  ,
\end{equation}
is given by
\begin{align}
\delta q &=  {1 \over T^2} {1 \over N}
\sum_{i, j, k}
{1 \over m} \sum_{\mu,\nu,\eta}
\Bigl[
\chi_{i j}^{\mu\nu}
\chi_{i k}^{\mu\eta} \Bigr]\av [\delta h_j^\nu \delta h_k^\eta]\av , \\
&= {1 \over T^2}\, \chi_{SG} \, \, \delta h_r^2 ,
\end{align}
where now
\begin{align}
\chi_{SG} &= {1 \over N}
\sum_{i, j} {1 \over m}
\sum_{\mu,\nu}
\Bigl[
(\chi_{i j}^{\mu\nu})^2 \Bigr]\av, \\
&= {1 \over N}
\sum_{i, j} {1 \over m}
\sum_{\mu,\nu}
\Bigl[
\left(\   \langle S_i^\mu S_j^\nu \rangle - \langle
S_i^\mu \rangle \langle S_j^\nu \rangle \ \right)^2  \Bigr]\av  .
\label{chisg}
\end{align}

For the Ising case, the sign of the field can be ``gauged away'' by the
transformation $S_i \to -S_i$, and $ J_{ij}  \to -J_{ij}$ for all $j$. Hence
the only difference between a uniform field and a Gaussian random field 
is that the latter varies in \textit{magnitude}, and these magnitude
fluctuations turn out to have only a minor effect\cite{bray:82b}. However, 
for the vector case, the random \textit{direction} of the
Gaussian random field does make a big difference because there is no longer a
distinction between longitudinal and transverse, and so there is \textit{no
longer a GT line to
preempt the AT line}.

In zero field, $\chi_{SG}$
diverges at the transition temperature $T_c$ given in Eq.~\eqref{Tc},
which is expected since $\chi_{SG}$ is the
susceptibility corresponding to the order parameter. Surprisingly, AT showed
for the Ising case ($m=1$)
that it also diverges in a magnetic field (either uniform, as
originally considered by
AT, or random, as considered later by Bray\cite{bray:82b}) along the AT line in the field-temperature
plane.
Below the AT line, $\chi_{SG}$ goes negative, indicating that
the RS solution is incorrect, and has to be replaced by the
Parisi\cite{parisi:79,parisi:80} 
replica symmetry breaking (RSB) solution.

In this paper we calculate $\chi_{SG}$ for a vector spin glass in the presence
of a random field, and show that it also becomes negative below an AT line in
the $h_r$--$T$ plane, whose location we calculate.
This fact does not appear to be widely recognized. Although
a field which is random in direction can presumably not be applied
experimentally, we feel that there is theoretical interest in our result
because a random field \textit{can} be applied in simulations.
Whether or not an AT line exists in finite-range spin glasses,
is a crucial difference between the
replica symmetry breaking (RSB)
picture\cite{parisi:79,parisi:80,parisi:83,mezard:87} of the spin glass state,
where it does occur, and the droplet picture
\cite{fisher:86,fisher:87,fisher:88,bray:86},  where it does not.
It has been found possible to simulate Heisenberg spin
glasses for significantly larger
sizes\cite{lee:07,viet:09,fernandez:09b} than Ising spin glasses, so our
results may give an additional avenue through which to investigate
numerically the nature
of the spin glass state.

The plan of this paper is as follows. In Sec.~\ref{sec:ising} we compute the
non-linear susceptibility for the Ising spin glass following the the
lines of AT. In  Sec.~\ref{sec:vector} we do the corresponding calculation for
the vector spin glass. This is followed in Sec.~\ref{sec:numerics} by a
numerical evaluation of the AT line for several values of $m$ and a
confirmation of the results by Monte Carlo simulations for the Heisenberg spin
glass, $m=3$. We summarize our results in Sec.~\ref{sec:conclusions}. Many of
the technical details are relegated to appendices.

\section{The spin glass susceptibility for Ising spin glasses}
\label{sec:ising}

In this section we review the calculation of the AT line for the Ising case.
In the next section we shall
use this approach
to derive the AT line for vector spin glasses.

The standard way of averaging in random systems is the replica trick, which
exploits the result
\begin{equation}
\ln Z = \lim_{n\to 0} {Z^n -1 \over n} \, .
\end{equation}
Applying this to the Ising ($m=1$) version of the Hamiltonian in
Eq.~\eqref{Ham}, one has
\begin{multline}
[Z^n]\av = \mathrm{Tr}\,
\exp \Bigl[
{(\beta J)^2 \over 2 N} \sum_{\langle i, j \rangle} \sum_{\alpha,\beta}
S_i^\alpha S_j^\alpha S_i^\beta S_j^\beta \\
+ {h_r^2 \over 2} \sum_i \sum_{\alpha,\beta} S_i^\alpha S_i^\beta
\Bigr] \, .
\label{Zna}
\end{multline}
We denote averages over the effective replica Hamiltonian in the exponential
on the RHS of Eq.~\eqref{Zna} by $\langle \cdots \rangle $.
Following standard
steps, see e.g.~Refs.~\onlinecite{sherrington:75,binder:86},
one obtains (omitting an unimportant overall
constant)
\begin{multline}
[Z^n]\av = \int_{-\infty}^\infty \Bigl(\prod_{(\alpha\beta)} \left({N \over 2
\pi}\right)^{1/2}\, (\beta J)\, dq_{\alpha\beta}\Bigr) \\
\times \exp\Bigl(-N {(\beta J)^2 \over 2} \sum_{(\alpha\beta)}
q_{\alpha\beta}^2 
\Bigr) 
\Bigl(\mathrm{Tr}\, \exp L[q_{\alpha\beta}]\Bigr)^N ,
\label{Zn}
\end{multline}
where 
$L[q_{\alpha\beta}]$ is given by
\begin{equation}
L[q_{\alpha\beta}] = \beta^2 \sum_{(\alpha\beta)} \left(J^2 q_{\alpha\beta} +
h_r^2\right)S^\alpha S^\beta \, ,
\label{Lq}
\end{equation}
the trace is over the spins $S^\alpha$, $\alpha = 1, \cdots,n$, and
$(\alpha\beta)$ denotes one of the $n(n-1)/2$ distinct pairs of replicas.

We take the replica symmetric (RS) saddle point,
where all the $q_{\alpha\beta}$
are equal to the same value $q$. The spin traces at the RS saddle point are evaluated
by writing
\begin{align}
\mathrm{Tr}\,e^L
&= \mathrm{Tr}\, \exp\Bigl(\beta^2 \sum_{(\alpha\beta)} \left(J^2 q +
h_r^2\right)S^\alpha S^\beta\Bigr)  \nonumber \\
&= \mathrm{Tr}\, 
\exp\left({\beta^2\over2} \left(J^2 q + h_r^2\right)
\bigl(\bigr\{\sum_\alpha
S^\alpha\bigr\}^2 - n\bigr)\right) \nonumber  \\
&\propto {1 \over \sqrt{2\pi}}
\int_{-\infty}^\infty e^{-z^2/2} \, d z \, \prod_{\alpha=1}^n \left[
\mathrm{Tr}\, e^{\beta(J^2 q + h_r^2)^{1/2} z S^\alpha }
\right]  ,
\label{Salpha_av}
\end{align}
where, in the last line, we omitted the constant factor
$\exp[-(\beta^2/2)(J^2 q + h_r^2) n]$, and decoupled the square in the
exponential using the identity
\begin{equation}
{1 \over \sqrt{2\pi}}
\int_{-\infty}^\infty e^{-z^2/2\, +\, a z} \, d z = e^{a^2/2} \, .
\label{HS}
\end{equation}

Consequently the replica spins $S^\alpha$ (without site label)
are independent of each other and feel a
Gaussian random field (the same for all replicas)
with zero mean and variance given by
\begin{equation}
\Delta^2 \equiv \beta^2 (J^2 q + h_r^2) \, .
\label{Delta2}
\end{equation}
We denote an average over the Gaussian random variable $z$ in
Eq.~\eqref{Salpha_av} by
$[\cdots]\z$, i.e.
\begin{equation}
[ f(z) ]\z = {1 \over \sqrt{2 \pi}} \int_{-\infty}^\infty
e^{-z^2/2} f(z) \, dz .
\label{fz}
\end{equation}
It is straightforward to evaluate averages over the $S^\alpha$, since they are
independent, so we will now express averages over the real spins $S_i$ in
terms of $S^\alpha$ averages.

One can show, see
e.g.~Ref.~\onlinecite{binder:86},
that each separate thermal average corresponds to a distinct replica,
so, for example,
\begin{equation}
[\langle S_i S_j \rangle  \langle S_k \rangle \langle S_l \rangle ]\av =
\langle S_i^\alpha S_j^\alpha S_k^\beta S_l^\gamma \rangle
\label{repl_av}
\end{equation}
for $\alpha, 
\beta$ and $\gamma$ all different.
To evaluate averages of the form in the RHS of Eq.~\eqref{repl_av}
we add fictitious fields $\Delta_{\alpha\beta}$ which couple the
replicas\cite{binder:86}, so Eq.~\eqref{Zna} becomes
\begin{multline}
[Z^n]\av = \mathrm{Tr}\, 
\exp \Biggl(
{(\beta J)^2 \over 2 N} \sum_{\langle i, j \rangle} \sum_{\alpha,\beta}
S_i^\alpha S_j^\alpha S_i^\beta S_j^\beta
\\
+ {h_r^2 \over 2} \sum_i \sum_{\alpha,\beta} S_i^\alpha S_i^\beta
+ \sum_{(\alpha\beta)} \Delta_{\alpha\beta} \sum_i S_i^\alpha S_i^\beta
\Biggr) \, .
\label{Znb}
\end{multline}
Taking derivatives with respect to $\Delta_{\alpha\beta}$, one has,
for $n \to 0$,
\begin{subequations}
\label{partialDelta}
\begin{align}
\sum_i \langle S_i^\alpha S_i^\beta \rangle &= {\partial \over \partial
\Delta_{\alpha\beta}}\, \left[Z^n \right]\av ,
\label{Sab}\\
\sum_{i, j} \langle S_i^\alpha S_i^\beta S_j^\gamma S_j^\delta
\rangle &= {\partial^2 \over \partial
\Delta_{\alpha\beta} \Delta_{\gamma\delta}}\, \left[Z^n \right]\av .
\label{Sabcd}
\end{align}
\end{subequations}
Now setting the $\Delta_{\alpha\beta}$ to zero
we get, from Eq.~\eqref{Znb}, in the $n\to 0$ limit,
\begin{equation}
q \equiv {1 \over N} [ \langle S_i\rangle^2 ]\av = 
{1 \over N} \sum_i \langle S_i^\alpha S_i^\beta \rangle = 
[ \langle S^\alpha S^\beta \rangle ]\z \, ,
\label{qSab}
\end{equation}
for $\alpha \ne \beta$.
We emphasize that, 
in the final average $[\langle ... \rangle ]\z$, the inner brackets refer to
averaging over the spins in a fixed value of the random field $z$ in
Eq.~\eqref{Salpha_av}, and the outer brackets, $[\cdots]\z$, refer to averaging over
$z$ according to Eq.~\eqref{fz}.
Equation \eqref{qSab} leads to
the well-known self-consistent expression\cite{sherrington:75,binder:86}
for the spin glass order parameter $q$:
\begin{align}
q &= [ \langle S^\alpha S^\beta \rangle ]\z \, , \nonumber \\
&= [ \tanh^2[ \beta \left(J^2 q + h_r^2\right)^{1/2} z]\z\, , \nonumber \\
&= 
{1\over \sqrt{2\pi}} \int_{-\infty}^\infty e^{-z^2/2}
\tanh^2[ \beta \left(J^2 q + h_r^2\right)^{1/2} z] \,\, d z \, .
\label{q_RS}
\end{align}

It will be useful to express the average in Eq.~\eqref{Sab} in a different way. 
Including the fictitious fields $\Delta_{\alpha\beta}$ in the derivation which
led from Eq.~\eqref{Zna} to Eqs.~\eqref{Zn} and \eqref{Lq} one finds an extra term,
$\sum_{(\alpha\beta)} \Delta_{\alpha\beta} S^\alpha S^\beta$, in
$L[q_{\alpha\beta}]$. Defining new integration variables by\cite{binder:86}
\begin{equation}
q_{\alpha\beta} + (\beta J)^{-2} \Delta_{\alpha\beta}
\rightarrow
q_{\alpha\beta} ,
\end{equation}
then $\Delta_{\alpha\beta}$ no longer appears in $L$, only in the quadratic
term in Eq.~\eqref{Zn}.  Using Eqs.~\eqref{partialDelta}, one then gets
\begin{subequations}
\label{partialDelta'}
\begin{align}
q = {1 \over N} \sum_i \langle S_i^\alpha S_i^\beta \rangle &= \langle q_{\alpha\beta}
\rangle \, ,
\label{Sab'}\\
{1 \over N} \sum_{i, j} \langle S_i^\alpha S_i^\beta S_j^\gamma S_j^\delta
\rangle &=
N \langle q_{\alpha\beta}\, q_{\gamma\delta} \rangle - (\beta J)^{-2} 
\delta_{(\alpha\beta),(\gamma\delta)} \, .
\label{Sabcd'}
\end{align}
\end{subequations}
Hence the spin glass susceptibility, defined in Eq.~\eqref{chisg_ising}, is given
by\cite{binder:86,missing}
\begin{equation}
\chi_{SG} = N\left( \langle \delta q_{\alpha\beta}^2 \rangle -
2 \langle \delta q_{\alpha\beta}\, \delta q_{\alpha\gamma} \rangle +
\langle \delta q_{\alpha\beta}\, \delta q_{\gamma\delta} \rangle \right) - (\beta
J)^{-2},
\label{chisg_qab}
\end{equation}
where all replicas are different, and $\delta q_{\alpha\beta}$ is defined by
\begin{equation}
q_{\alpha\beta} =
q + \delta q_{\alpha\beta} .
\end{equation}

We now expand Eq.~\eqref{Zn}
about the saddle point
to quadratic order in the $\delta q_{\alpha\beta}$.
The result is that the exponential in Eq.~\eqref{Zn} becomes
\begin{equation}
\exp\Bigl(-N f(q) -N  {(\beta J)^2 \over 2}
\sum_{(\alpha\beta), (\gamma\delta)} A_{(\alpha\beta),(\gamma\delta)}
\delta q_{\alpha\beta} \delta q_{\gamma\delta}
\Bigr)  ,
\label{qab_qcd}
\end{equation}
where $f(q)$ is the value of the exponent
at the saddle point. To obtain the elements of the $\smfrac{1}{2}n(n-1)$
by $\smfrac{1}{2}n(n-1)$ matrix $A$ we take the log of Eq.~\eqref{Zn} and write
the coefficients in the expansion of $\ln \mathrm{Tr}\, e^L$ in powers of
the $\delta q_{\alpha\beta}$ in terms of spin averages, evaluated by the
decoupling in Eq.~\eqref{Salpha_av}. The result is
\begin{multline}
A_{(\alpha\beta),(\gamma\delta)} = \delta_{(\alpha\beta)(\gamma\delta)} - \\
\left(\beta J\right)^2 \left\{ \left[ \langle S^\alpha S^\beta S^\gamma S^\delta
\rangle\right]\z - \left[\langle  S^\alpha S^\beta \rangle\right]\z \,
\left[\langle S^\gamma S^\delta
\rangle \right]\z \right\} .
\label{A-Isingb}
\end{multline}

Equation \eqref{qab_qcd} is the weight function used for averaging over the 
$\delta q_{\alpha\beta}$ in Eq.~\eqref{chisg_qab}. Performing
these Gaussian integrals
gives
\begin{multline}
\chi_{SG} = {1 \over (\beta J)^2} \, \Bigl[
G_{(\alpha\beta),(\alpha\beta)} -
2 G_{(\alpha\beta),(\alpha\gamma)} + \\
G_{(\alpha\beta),(\gamma\delta)} - 1 \Bigr] \, ,
\label{chisg-qabcd}
\end{multline}
where $G$ is the matrix inverse of $A$, i.e.
\begin{equation}
G\, A = I
\end{equation}
where $I$ is the identity matrix.
Defining
\begin{subequations}
\begin{align}
G_{(\alpha\beta),(\alpha\beta)} & = G_1 , \\
G_{(\alpha\beta),(\alpha\gamma)} & = G_2 , \\
G_{(\alpha\beta),(\gamma\delta)} & = G_3 , 
\end{align}
\label{G123}
\end{subequations}
we have
\begin{equation}
\chi_{SG} = {1 \over (\beta J)^2} \, \left(G_r - 1 \right) \, ,
\label{chisg-Gr}
\end{equation}
where
\begin{equation}
G_r = G_1 - 2 G_2 + G_3
\end{equation}
is called the ``replicon propagator''\cite{bray:79c}.

The matrix inverse of $A$
is evaluated in
Appendix
\ref{sec:inverse}.
According to Eq.~\eqref{Gr} we can express
Eq.~\eqref{chisg-Gr} as
\begin{equation}
\chi_{SG} = {1 \over (\beta J)^2}\, \left({1 \over \lambda_3} - 1\right) ,
\label{chi-la3}
\end{equation}
where 
\begin{equation}
\lambda_3 = P -2 Q + R,
\end{equation}
and the quantities
$P, Q$ and $R$ are defined in Eq.~\eqref{A}.
The eigenvalues of $A$ are evaluated in
Appendices \ref{sec:l1}--\ref{sec:l3}, and it turns out that
$\lambda_3$ is an eigenvalue of $A$, see Eq.~\eqref{lambda3}.
We evaluate the relevant spin averages needed to determine
$\lambda_3$ in in Appendix \ref{sec:averages}, and
Eq.~\eqref{lambda3S0}
gives
\begin{equation}
\lambda_3 = 1 - (\beta J)^2 \chi_{SG}^0\, ,
\end{equation}
or equivalently, from Eq.~\eqref{chi-la3},
\begin{equation}
\chi_{SG} = {\chi_{SG}^0 \over 1 - (\beta J)^2 \chi_{SG}^0} ,
\label{chisg_RPA}
\end{equation}
where $\chi_{SG}^0$ is a single-site spin glass susceptibility, given for the
Ising case by
\begin{align}
\chi_{SG}^0 &=  
\left[ \Bigl(
\langle S S \rangle - 
\langle S \rangle \langle S \rangle \Bigr)^2
\right]\z , \nonumber   \\
&= \left[ \Bigl(1 - \langle S \rangle^2 \Bigr)^2 \right]\z , \nonumber \\
&= 
\left[ \Bigl(1 - 
\tanh^2[ \beta \left(J^2 q + h_r^2\right)^{1/2} z] 
\Bigr)^2 \right]\z , \nonumber \\
&= 1 - 2 q + r ,
\label{chisg0_Ising}
\end{align}
where $q$ is given by Eq.~\eqref{q_RS} and $r$ is given by
\begin{equation}
r = 
{1\over \sqrt{2\pi}} \int_{-\infty}^\infty e^{-z^2/2}
\tanh^4[ \beta \left(J^2 q + h_r^2\right)^{1/2} z] \,\, d z \, .
\end{equation}
Hence, according to the RS ansatz, $\chi_{SG}$ is predicted
to diverge
where
\begin{equation}
(\beta J)^2 \chi_{SG}^0 = 1,
\end{equation}
which describes the location of the AT line. In particular, for small fields
the AT line is given by
\begin{equation}
h_r^2 = {4 \over 3} \left({T_c - T \over T_c}\right)^3  \quad (m=1)\, ,
\end{equation}
see Eq.~\eqref{HrAT}.
In fact, $\chi_{SG}$
turns out to be negative below this line since $\lambda_3$ is 
negative
in this region, see Eq.~\eqref{unstable}. These 
results were first found by AT.
At low temperatures we get
\begin{equation}
{h_r(T\to0) \over J}  = \sqrt{8 \over 9 \pi} \,\, 
{J \over T} \quad (m=1) \, ,
\end{equation}
see Eq.~\eqref{lowTising},
in agreement with Bray\cite{bray:82b}.
A plot of the AT line for $m=1$,
obtained numerically, is shown in Fig.~\ref{fig:ATline}.

Although the derivation of Eq.~\eqref{chisg_RPA} is rather involved, we note that the
final answer is quite simple and has a familiar mean field form, i.e.~a
response function $\chi$ is equal to $\chi_0 /( 1 - K \chi_0)$ where $\chi_0$ is
the non-interacting response function, and $K$ ($= (\beta J)^2$ here),
is a coupling constant. In the
next section, we will see that $\chi_{SG}$ has precisely the same mean field form
for the vector ($m>1$) case.

%
%

\section{The spin glass susceptibility for vector spin glasses}
\label{sec:vector}

Here we consider a vector spin glass in which the Ising spins are replaced by vector
spins with $m$ components. The fluctuations in zero field were first
considered by AJKT and Ref.~\onlinecite{almeida:80} and our approach follows
closely that of the latter reference. However, we shall see that there are
some differences between our results and those of AJKT and
Ref.~\onlinecite{almeida:80}.
The derivation follows the
lines of that for the Ising case in the previous section, but with the 
burden of additional
indices for the spin components. Hence we will not go through the details
but just indicate the main steps and the results.

To avoid
confusion in notation, we will use the Greek
letters $\alpha,\beta,\gamma,\delta,\epsilon$ for replicas and
$\mu,\nu,\kappa,\sigma$
for spin indices. The auxiliary variables $q$ will now involve 4
indices $(\alpha\beta), \mu\nu$, in which the order of the replica pair
$(\alpha\beta)$ is unimportant, i.e.~$(\beta\alpha)$
is the same as $(\alpha\beta)$,
but the order of the spin indices does matter because
$S_\alpha^\mu S_\beta^\nu$ is not the same as $S_\alpha^\nu S_\beta^\mu$.
Another new feature which appears when we deal with vector spins is the
appearance of terms with both replicas equal, $(\alpha\alpha)$.  These do not
appear for the Ising case because $\left(S_\alpha\right)^2$ is equal to 1, a
constant. However, $\left(S_\alpha^\mu\right)^2$ is not a constant for $m > 1$
and so we now need to include $(\alpha\alpha)$ terms
in the analysis, though they will not enter the final result for
$\chi_{SG}$.

The analogues of Eqs.~\eqref{Zn} and \eqref{Lq} are
\begin{widetext}
\begin{multline}
[Z^n]\av = \int_{-\infty}^\infty
\Bigl(\prod_{(\alpha\beta),\mu,\nu} \left({N \over 2 \pi}\right)^{1/2}\,
(\beta J)\, dq_{\alpha\beta}^{\mu\nu}\Bigr) \,
\Bigl(\prod_{\alpha,\mu,\nu} \left({N \over 2 \pi}\right)^{1/2}\,
(\beta J)\, dq_{\alpha\alpha}^{\mu\nu}\Bigr) \\
\times \exp\biggl(-N {(\beta J)^2 \over 2} \Bigl\{\sum_{(\alpha\beta),\mu,\nu}
\left(q_{\alpha\beta}^{\mu\nu}\right)^2 
+ \sum_{\alpha,\mu,\nu} \left(q_{\alpha\alpha}^{\mu\nu}\right)^2  \Bigr\} 
\biggr) 
\Bigl( \mathrm{Tr}\, \exp L[q_{\alpha\beta}^{\mu\nu},q_{\alpha\alpha}^{\mu\nu}]
\Bigr)^N \, ,
\label{Zn-vector}
\end{multline}
where
$L[q_{\alpha\beta}^{\mu\nu}, q_{\alpha\alpha}^{\mu\nu}]$ is given by
\begin{equation}
L[q_{\alpha\beta}^{\mu\nu},q_{\alpha\alpha}^{\mu\nu}]
= \beta^2 \sum_{(\alpha\beta),\mu,\nu} \left(J^2 q_{\alpha\beta}^{\mu\nu} +
h_r^2 \delta_{\mu\nu} \right)S^\alpha_\mu S^\beta_\nu +
{(\beta J) ^2 \over \sqrt{2}} \sum_{\alpha,\mu,\nu} q_{\alpha\alpha}^{\mu\nu} 
S^\alpha_\mu S^\alpha_\nu \, ,
\label{Lq-vector}
\end{equation}
\end{widetext}
where we ignored a term $\smfrac{1}{2}(\beta h_r)^2 \sum_{\mu,\alpha}
\left(S_\mu^\alpha\right)^2$
since it is a constant.

We take the replica symmetric (RS) saddle point,
where
\begin{equation}
q_{\alpha\beta}^{\mu\nu} = q\, \delta_{\mu\nu}, \quad q_{\alpha\alpha}^{\mu\nu}
= x\, \delta_{\mu\nu}\, .
\end{equation}
We then have, ignoring overall constant factors,
\begin{widetext}
\begin{align}
e^L
&\propto \mathrm{Tr}\, \exp\Bigl(\beta^2 \sum_{(\alpha\beta),\mu,\nu} \left(J^2 q +
h_r^2\right)S^\alpha_\mu S^\beta_\mu \Bigr)  
= \mathrm{Tr}\, 
\exp\left({\beta^2\over2} \left(J^2 q + h_r^2\right) \sum_\mu
\bigl(\bigr\{\sum_\alpha
S^\alpha_\mu \bigr\}^2\bigr) - n m\right) \nonumber  \\
&\propto 
\int_{-\infty}^\infty \left( \prod_\mu {dz_\mu \over \sqrt{2 \pi}} \right) 
e^{-\sum_\mu z_\mu^2/2} \, \prod_{\alpha=1}^n \left[
\mathrm{Tr}\, e^{\beta(J^2 q + h_r^2)^{1/2} \sum_\mu z_\mu S^\alpha_\mu }
\right]  ,
\label{Salpha_av-vector}
\end{align}
\end{widetext}
where, to get the last line, we decoupled the square in the exponent using
Eq.~\eqref{HS}.
As for the Ising case, we denote an average over the Gaussian random
variables $z_\mu$ by $[\cdots]\z$.

Proceeding as in Sec.~\ref{sec:ising}, the spin glass susceptibility, defined
in Eq.~\eqref{chisg}, is given by
\begin{widetext}
\begin{equation}
\chi_{SG} = {N \over m} \Biggl(\sum_{\mu,\nu}
\langle \delta q_{\alpha\beta}^{\mu\mu} \,
\delta q_{\alpha\beta}^{\nu\nu}\rangle -
2 \langle \delta q_{\alpha\beta}^{\mu\mu}\, 
\delta q_{\alpha\gamma}^{\nu\nu} \rangle
+ \langle \delta q_{\alpha\beta}^{\mu\mu}\, 
\delta q_{\gamma\delta}^{\nu\nu} \rangle \Biggr)
- (\beta J)^{-2},
\label{chisg_qab-vector}
\end{equation}
(with $\alpha, \beta, \gamma$ and $\delta$ all different)
where the averages over the $\delta q$ are with respect to the following
Gaussian weight (analogous to that in Eq.~\eqref{qab_qcd} for the Ising case),
\begin{equation}
\exp\Biggl(-N {(\beta J)^2 \over 2} \Bigl\{
\sum_{(\alpha\beta), (\gamma\delta)}
Z_{(\alpha\beta),(\gamma\delta)}^{\mu\nu,\kappa\sigma}
\delta q_{\alpha\beta}^{\mu\nu} \delta q_{\gamma\delta}^{\kappa\sigma} + 
\sum_{\alpha, (\gamma\delta)}
Z_{(\alpha\alpha),(\gamma\delta)}^{\mu\nu,\kappa\sigma}
{\delta q_{\alpha\alpha}^{\mu\nu} \over \sqrt{2}} \delta q_{\gamma\delta}^{\kappa\sigma} + 
\sum_{\alpha, \gamma}
Z_{(\alpha\alpha),(\gamma\gamma)}^{\mu\nu,\kappa\sigma}
{\delta q_{\alpha\alpha}^{\mu\nu} \over\sqrt{2}}
{\delta q_{\gamma\gamma}^{\kappa\sigma} \over\sqrt{2}}
\bigr\}
\Biggr)  ,
\label{qab_qcd-vector}
\end{equation}
\end{widetext}
and
\begin{multline}
Z_{(\alpha\beta),(\gamma\delta)}^{\mu\nu,\kappa\sigma}
= \delta_{(\alpha\beta)(\gamma\delta)} \delta_{\mu\kappa} \delta_{\nu\sigma} - \\
\left(\beta J\right)^2 \left\{
\left[\langle S^\alpha_\mu S^\beta_\nu S^\gamma_\kappa S^\delta_\sigma
\rangle\right]\z - \left[\langle  S^\alpha_\mu S^\beta_\nu \rangle \right]\,
\left[\langle S^\gamma_\kappa S^\delta_\sigma
\rangle \right]\z
\right\}.
\label{Z-vectorb}
\end{multline}
Note that the annoying factors of $1/\sqrt{2}$ and $1/2$ in
Eq.~\eqref{qab_qcd-vector} can
be removed simply by incorporating a factor of $1/\sqrt{2}$ into the 
$q_{\alpha\alpha}^{\mu\nu}$. Doing the averages in Eq.~\eqref{chisg_qab-vector}
using the
Gaussian weight in Eq.~\eqref{qab_qcd-vector} gives
\begin{multline}
\chi_{SG} = {1 \over (\beta J)^2}  \, \Bigl\{ {1 \over m} \sum_{\mu,\nu} \Bigl[
G_{(\alpha\beta),(\alpha\beta)}^{\mu\mu\nu\nu} -
2 G_{(\alpha\beta),(\alpha\gamma)}^{\mu\mu\nu\nu} + \\
G_{(\alpha\beta),(\gamma\delta)}^{\mu\mu\nu\nu}\Bigr] - 1 \Bigr\} \, ,
\label{chisg-vector-qabcd}
\end{multline}
where $G = Z^{-1}$. Using the definitions in Eqs.~\eqref{G123LT}, we have
\begin{equation}
\chi_{SG} = {1 \over (\beta J)^2} \, (G_r - 1)
\label{chisg-Gr-vector}
\end{equation}
where the ``replicon'' propagator is given by
\begin{multline}
G_r = G_{1L} + (m-1)G_{1T} - 2 \left[G_{2L} + (m-1)G_{2T}\right] + \\
G_{3L} + (m-1)G_{3T} \, .
\label{Grb}
\end{multline}
The matrix inverse of $Z$ is evaluated in Appendix \ref{sec:inverse-vector}.
According to
Eq.~\eqref{G-vector}, we can express
Eq.~\eqref{chisg-Gr-vector} as
\begin{equation}
\chi_{SG} = {1 \over (\beta J)^2} \, \left({1 \over \lambda_{3S}} - 1\right),
\label{chisg_l3}
\end{equation}
where
\begin{multline}
\lambda_{3S} = P_L + (m-1)P_T - 2 \left[Q_L + (m-1)Q_T\right]
+ \\
R_L + (m-1)R_T \, .
\end{multline}
We determine the eigenvalues of $Z$ in
Appendices \ref{sec:subspaces}--\ref{sec:-}, and
show that $\lambda_{3S}$ is an eigenvalue,

From Eq.~\eqref{lambda3S0}, we see that Eq.~\eqref{chisg-Gr-vector}
can be written in the same form
as for the Ising case, Eq.~\eqref{chisg_RPA}, 
where,
for the case of general $m$, the
single-site spin glass susceptibility $\chi_{SG}^0$ is evaluated in Appendix
\ref{sec:averages}, and given by
Eq.~\eqref{l3s_m}.

The AT line is where $(\beta J)^2 \chi_{SG}^0 = 1$.
Near $T_c$ this is given by
\begin{equation}
\left( {h_r \over J}\right) ^2 = {4 \over m + 2}\, t^3 ,
\label{HrATb}
\end{equation}
see Eq.~\eqref{HrAT}.
The same expression was obtained by Gabay and
Toulouse\cite{gabay:81} but for a uniform field, in which case it refers to a
crossover rather than a sharp transition.
Note that $h_r = 0$ for $m
= \infty$, as expected since AJKT showed that 
the replica symmetic solution is stable in this
limit.
In the opposite limit, $T \to 0$, we find
that the value of the AT field is finite for $m > 2$,
\begin{equation}
{h_r(T = 0) \over J} = {1 \over \sqrt{m-2}} \quad (m > 2) \, ,
\label{lowTmb}
\end{equation}
see Eq.~\eqref{lowTm},
while $h_r(T \to 0)$ diverges for $m \le 2$. 
The location of the AT line, obtained numerically,
is plotted in Fig.~\ref{fig:ATline} for several values of $m$.

Below the AT line, $\chi_{SG}$ is
predicted to be negative, see
Eq.~\eqref{unstable}, which is impossible and shows that the RS solution (which we
have assumed) is wrong in this region.

For $h_r = 0$, Eq.~\eqref{unstable} gives $\lambda_{3S} = - 4t^2 /( m +
2)$, which disagrees with the unstable eigenvalue 
$-8 t^2 / (m+2)^2$ given by AJKT and Ref.~\onlinecite{almeida:80}.
However, we note that the 
replicon propagator in
Eq.~\eqref{Grb} corresponds precisely to Eq.~(3.5) of
Ref.~\onlinecite{bray:79b}, and Eq.~\eqref{HrATb} also appears in the paper by
Gabay and Toulouse\cite{gabay:81}, so
we are confident that Eq.~\eqref{unstable} is correct. 
Note too that we obtained the spin glass susceptibility, the divergence of which
indicates the AT line, \textit{directly}
from the inverse of the matrix $Z$, the calculation of which is fairly simple,
see Appendix \ref{sec:inverse-vector}.
The extra information that
$\chi_{SG}$ is related to
an eigenvalue, $\lambda_{3S}$, is not strictly needed to locate 
the AT line. 

\section{Numerical results}
\label{sec:numerics}

\begin{figure}
\includegraphics[width=\figurewidth]{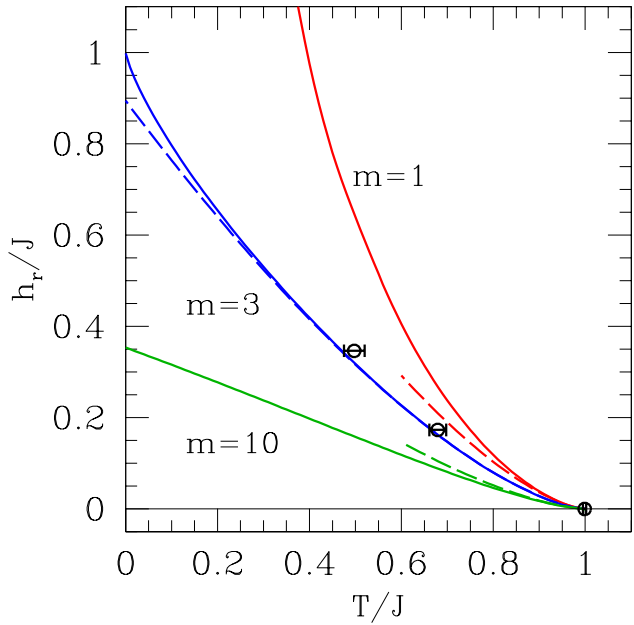}
\caption{
The solid lines indicate the location of the AT line for $m=1, 3$ and $10$,
according to Eq.~\eqref{final_res}, and $\chi_{SG}^0$ given by
Eq.~\eqref{l3s_m}. For $m \to\infty$ the AT line collapses on to the
horizontal axis. The dashed lines are the approximate form given in
Eq.~\eqref{HrATb}, which is valid close to $T= T_c = J$. Note that
this approximation works
remarkably well for the Heisenberg case, $m=3$, even down to quite low
temperatures. Also shown are Monte Carlo results for the critical temperature
for $h_r = 0, 0.173$ and $0.346$ for $m=3$.
\label{fig:ATline}
}
\end{figure}

We have determined the location of the AT line numerically for $m=1, 3$ and
$10$. For a given $T$ and assumed value of $h_r$ we solve for $q$
self-consistently from Eq.~\eqref{q} and substitute into Eq.~\eqref{l3s_m} which
gives $\lambda_{3S}$ from Eq.~\eqref{lambda3S0}.
The value of $h_r$ is then adjusted until $\lambda_{3S}
= 0$. The results are shown by the solid lines in Fig.~\ref{fig:ATline}. Also
shown, by the dashed lines, is the approximate form in Eq.~\eqref{HrATb} which is
valid close to the zero field transition temperature. For $m=3$ this
approximation actually
works well down to rather low temperatures.

If the spins are normalized to have length 1 rather than $m^{1/2}$ one divides
the horizontal scale in Fig.~\ref{fig:ATline} by $m$ and the vertical scale by
$1/m^{1/2}$, so the zero field transition
temperature would be $T_c = J/m$ and the zero temperature limit of the AT field would be
$h_r = J / \sqrt{m(m-2)}$, for $m>2$ (compare with Eq.~\eqref{lowTmb}).

We have also checked these results by Monte Carlo simulations for the
Heisenberg case, $m=3$. The method has been discussed
elsewhere\cite{lee:07,fernandez:09b}, so here we just give a few salient
features. We use three types of moves: heatbath, overrelaxation, and 
parallel tempering\cite{hukushima:96,marinari:98b}. We perform one heatbath
sweep and one parallel tempering sweep for every ten overrelaxation sweeps. The
parameters of the simulations are given in Table \ref{tab:params}. In
calculating the spin glass susceptibility in Eq.~\eqref{chisg}, each thermal average
is run in a separate copy of the system to avoid bias. Hence we simulate four
copies at each temperature.

When the quenched random disorder variables are Gaussian, as here, the
following identity is easily shown to hold by integrating by parts the
expression for the average energy $U$ with
respect to the disorder variables\cite{katzgraber:05,lee:07},
\begin{equation}
-{U \over m} \equiv {[ \langle \mathcal{H}\rangle ]\av  \over m}
= {J^2 \over 2 T}\, (q_s - q_l) +
{h_r^2 \over T} \, (1 - \overline{q}),
\label{equil}
\end{equation}
where
\begin{align}
q_s &= {1 \over N m} \sum_{i\ne j} [\langle\, \bigl(
\mathbf{S}_i \cdot \mathbf{S}_j\bigr)^2 \rangle]\av\, ,\\
q_l &= {1 \over N m } \sum_{i\ne j} \Bigl[
\langle \mathbf{S}_i \cdot \mathbf{S}_j \rangle^2 \Bigr]\av\, ,\\
\overline{q} &=  {1 \over N m} \sum_{i} [
\langle \mathbf{S}_i \rangle \cdot
\langle \mathbf{S}_i \rangle
]\av \, ,
\end{align}
in which $\overline{q}$ is the expectation value of the spin glass order
parameter, and $q_l$ is called the ``link'' overlap.

While Eq.~\eqref{equil} is true in
equilibrium, is not true before equilibrium is reached, and, indeed, the two
sides of the equation approach the equilibrium value from opposite
directions\cite{katzgraber:05,lee:07}. Hence we only accept the results of
a simulation if Eq.~\eqref{equil} is satisfied with small error bars. (Note
that this equation refers to an average over samples; the connection between
the energy and the spin correlations 
is not true sample by sample.)

\begin{table}[!tb]
\caption{
Parameters of the simulations for different values of $hr$. Here
$N_{\rm samp}$ is the number of samples, $N_{\rm sweep}$ is the
number of overrelaxation Monte Carlo sweeps,
$T_{\rm min}$ and $T_{\rm max}$
are the lowest and highest temperatures simulated, and $N_T$ is the
number of temperatures.
\label{tab:params}}
\begin{tabular*}{\columnwidth}{@{\extracolsep{\fill}} |r| r r r r r l|}
\hline
\hline
$h_r$  &  $N$ & $N_\text{samp}$ & $N_\text{sweep}$ & $T_\text{min}$ & $T_\text{max}$ & $N_{T}$ 
\\
\hline
0     &   64 & 8000 &   256 & 0.30 & 1.50 & 40 \\
0     &  128 & 8000 &   512 & 0.30 & 1.50 & 40 \\
0     &  256 & 8000 &  1024 & 0.30 & 1.50 & 40 \\
0     &  512 & 8000 &  2048 & 0.30 & 1.50 & 40 \\
0     & 1024 & 2078 &  4096 & 0.30 & 1.50 & 40 \\[2mm]
\hline
0.173 &   64 & 8000 &  1024 & 0.30 & 1.50 & 40 \\
0.173 &  128 & 8000 &  2048 & 0.30 & 1.50 & 40 \\
0.173 &  256 & 8000 &  4096 & 0.30 & 1.50 & 40 \\
0.173 &  512 & 4279 &  8192 & 0.30 & 1.50 & 40 \\
0.173 & 1024 & 1494 & 16384 & 0.39 & 1.50 & 40 \\[2mm]
\hline
0.346 &   64 & 8000 &  1024 & 0.15 & 1.20 & 40 \\
0.346 &  128 & 8000 &  2048 & 0.15 & 1.20 & 40 \\
0.346 &  256 & 8000 &  4096 & 0.15 & 1.20 & 40 \\
0.346 &  512 & 4293 &  8192 & 0.15 & 1.20 & 40 \\
0.346 & 1024 & 3037 & 16384 & 0.15 & 1.20 & 40 \\
\hline
\hline
\end{tabular*}
\end{table}

\begin{figure}
\includegraphics[width=\figurewidth]{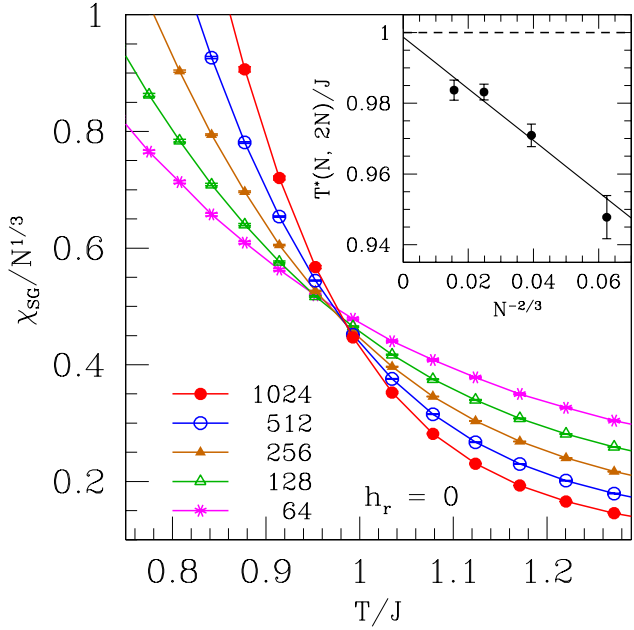}
\caption{
Zero field Monte Carlo data for the spin glass susceptibility for the $m=3$
(Heisenberg) infinite-range spin glass,
divided by $N^{1/3}$,
for different sizes.
According to finite-size scaling, the data should intersect at the
transition temperature $T_c$ in the absence of corrections to scaling.
Allowing for the leading corrections,
the inset shows intersection temperatures
$T^\star(N, 2N)$ for sizes $N$ and $2N$ and the extrapolation 
to $N = \infty$ according to Eq.~\eqref{Tstar}.  This leads to the estimate
$T_c = 0.9987 \pm 0.0036$ (see Table \ref{tab:Tstar}),
which agrees well with the exact value of
1, shown as the dashed line in the inset.
\label{fig:fig_0}
}
\end{figure}
\begin{figure}
\includegraphics[width=\figurewidth]{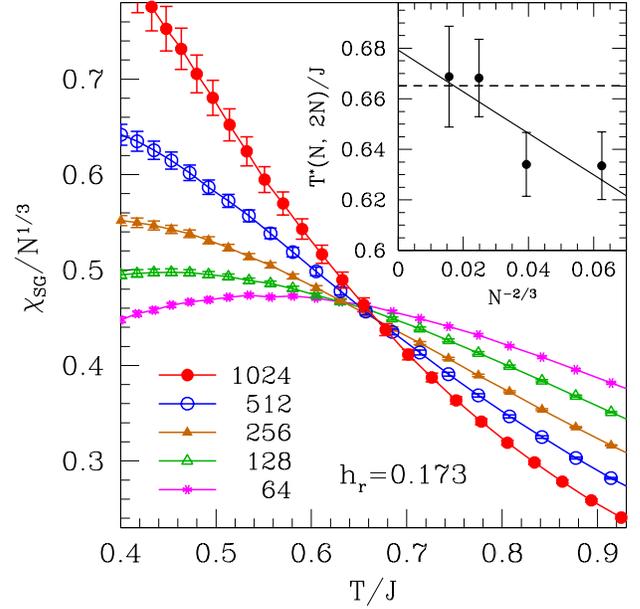}
\caption{
Same as Fig.~\ref{fig:fig_0} but for
random field strength $h_r = 0.173$. The final estimate of $T_c(h_r)$ is
$0.685 \pm 0.019$ which is to be compared with the exact value of $0.6652$,
see Table \ref{tab:Tstar}, which is shown as the dashed line in the inset.
\label{fig:fig_0.1}
}
\end{figure}
\begin{figure}
\includegraphics[width=\figurewidth]{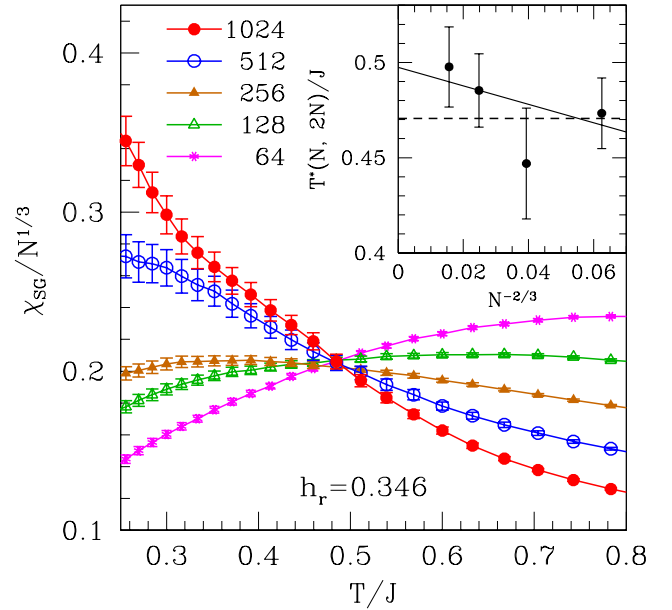}
\caption{
Same as Fig.~\ref{fig:fig_0} but for
random field strength $h_r = 0.346$. The final estimate of $T_c(h_r)$ is
$0.497 \pm 0.023$, to be compared with the exact value of $0.4706$,
see Table \ref{tab:Tstar},
which is shown as the dashed line in the inset.
\label{fig:fig_0.2}
}
\end{figure}

According to finite-size scaling the spin glass susceptibility in a finite,
infinite-range
system, should vary as\cite{billoire:03b,jorg:08b,larson:10,takahashi:10}
\begin{equation}
\chi_{SG} = N^{1/3} \widetilde{X}\left(N^{1/3} (T - T_c(h_r))\right) ,
\label{fss}
\end{equation}
so plots of $\chi_{SG}/N^{1/3}$ should intersect at the transition
temperature $T_c(h_r)$.
Data for $\chi_{SG} / N^{1/3}$ for $m=3$ for random field values
$h_r = 0$, 0.173 and 0.346,
are shown in
Figs.~\ref{fig:fig_0}, \ref{fig:fig_0.1} and \ref{fig:fig_0.2}.
The data does indeed intersect, indicating a
transition, though the data for different sizes don't intersect at exactly
the same temperature which indicates the presence of
corrections to finite-size scaling.

There are both singular and analytic corrections to scaling.
In the mean field limit the leading correction to $\chi_{SG}$ is
analytic\cite{fss-corrections}, in
fact just a constant, so we replace Eq.~\eqref{fss} by
\begin{equation}
\chi_{SG} = N^{1/3} \widetilde{X}\left(N^{1/3} (T - T_c(h_r))\right) + c_0 .
\label{fss2}
\end{equation}
We compute the intersection temperature $T^\star(N, 2N)$ between data for
$\chi_{SG}/N^{1/3}$ for sizes $N$ and $2 N$. It is then easy to see from
Eq.~\eqref{fss2} that
the $T^\star(N, 2N)$ converge to the transition temperature like
\begin{equation}
T^\star(N, 2N) - T_c(h_r) = {A \over N^{2/3} } \, ,
\label{Tstar}
\end{equation}
where the constant $A$ is related to $c_0$ and $\widetilde{X}'(0)$. We determine
$T^\star(N, 2N)$ by a bootstrap analysis and show the results both in Table
\ref{tab:Tstar} and in the insets to Figs.~\ref{fig:fig_0},
\ref{fig:fig_0.1} and \ref{fig:fig_0.2}.
Fitting a straight line through 
$T^\star(N, 2N)$ against $N^{-2/3}$ according to
Eq.~\eqref{Tstar}, gives estimates of $T_c$ which shown both in
Fig.~\ref{fig:ATline}
and Table \ref{tab:Tstar}.

\begin{table}[!tb]
\caption{
Intersection temperatures $T^\star(N, 2N)$, and extrapolated values of
$T_c(h_r)$ determined from fits to Eq.~\eqref{Tstar}. Also shown is the exact
value for $T_c(h_r)$,
obtained as described in the text.
\label{tab:Tstar}}
\begin{tabular*}{\columnwidth}{@{\extracolsep{\fill}}| l l | l | l | l |}
\hline
\hline
$h_r$  &  $N$ & $T^\star(N, 2N)$   & $T_c(h_r)$ & $T_c(h_r)$ (exact)
\\
\hline
0     &   64   & 0.9478(61) & & \\
0     &  128   & 0.9709(32) & & \\
0     &  256   & 0.9832(22) & & \\
0     &  512   & 0.9837(29) & & \\
\hline
0     &$\infty$&            & 0.9987(36) & 1 \\
\hline 
\hline 
0.173 &   64   & 0.633(13) & & \\
0.173 &  128   & 0.634(13) & & \\
0.173 &  256   & 0.668(15) & & \\
0.173 &  512   & 0.680(18) & & \\
\hline 
0.173 &$\infty$&           & 0.679(19) & 0.6652 \\
\hline 
\hline 
0.346 &   64   & 0.473(19) & & \\
0.346 &  128   & 0.447(29) & & \\
0.346 &  256   & 0.485(19) & & \\
0.346 &  512   & 0.498(21) & & \\
\hline
0.346 &$\infty$&            & 0.497(23) & 0.4706 \\
\hline
\hline
\end{tabular*}
\end{table}

We see that, in zero field, the numerics accurately gives the exact value for
$T_c$ of
1, and for non-zero $h_r$, the numerics gives the correct
answer to within about one sigma.
Hence our analytical predictions for the AT line in Heisenberg spin glasses
are well confirmed by simulations.

\section{Conclusions}
\label{sec:conclusions}

We have emphasized that the appropriate symmetry breaking field for a spin
glass is a random field, and that,
for a vector spin glass, the crucial ingredient is the random 
\textit{direction} of the field. Incorporating a random field, there is a line of
transitions (AT line) in vector spin glasses, just as there is in the Ising spin
glass, a fact which does not seem to be widely recognized.
The AT line is
different from the Gabay-Toulouse\cite{gabay:81} (GT) line, since it is a
transition to a phase with replica symmetry breaking but \textit{no change in
spin symmetry}.

The location of the AT line for vector spin glasses with Gaussian random fields
is given by
\begin{equation}
\left({T \over J}\right)^2 = \chi_{SG}^0 ,
\label{final_res}
\end{equation}
where $\chi_{SG}^0$ is given by
Eq.~\eqref{l3s_m}. 
For the important case of the Heisenberg ($m=3$)
spin glass,
the simpler expression for $\chi_{SG}^0$
is given in Eq.~\eqref{l3s_Heis}.
We have plotted the AT line numerically for several values of $m$ in
Fig.~\ref{fig:ATline},
and confirmed these results numerically by simulations for the case
of $m = 3$.

For the Ising case, 
we note that Bray and Moore\cite{bray:79} have obtained Eq.~\eqref{chisg_RPA}
for the spin glass susceptibility
without
replicas, starting
starting from the local mean-field equations of Thouless, Anderson and
Palmer~\cite{thouless:77} (the TAP equations). It would be interesting to see
if one could derive, along similar lines,
a more straightforward, and non-replica, calculation of
$\chi_{SG}$ for the vector spin case too. 

Although it is not possible experimentally
to apply a field which is random in direction to a vector spin
glass, so the AT line seems to be experimentally inaccessible (except for
the Ising case), one
\textit{can} detect the AT line for vector spin glasses
in simulations. 
Whether or not at AT line exists in finite-range spin glasses,
is a crucial difference between the replica symmetry breaking (RSB) picture,
where it does occur, and the droplet picture, where it does not.
It has been found possible to simulate Heisenberg spin
glasses for significantly larger
sizes\cite{lee:07,viet:09,fernandez:09b} than Ising spin glasses, so our
results may give an additional avenue through which to investigate the nature
of the spin glass state.


\begin{acknowledgments}
This work is supported by the National Science Foundation under Grant
No.~DMR-0906366. We are grateful for a generous
allocation of computer time from the Hierarchical Systems Research
Foundation, and are particularly grateful to Jairo de Almeida for making
available to us the relevant sections of his thesis,
Ref.~\onlinecite{almeida:80}, and for comments on an earlier version of this
manuscript.
\end{acknowledgments}

\appendix

\section{Fluctuation analysis for Ising spin glasses}
\label{sec:diag-Ising}

We follow AT in
obtaining the eigenvalues and eigenvectors (and also the inverse, not
calculated by AT) of
the real symmetric
matrix $A$ of dimension $n(n-1)/2$, in which each row of column is labeled by
a pair of distinct spin indices $(\alpha\beta)$, with elements given by (see
Eq.~\eqref{A-Isingb})
\begin{multline}
A_{(\alpha\beta),(\gamma\delta)} = \delta_{(\alpha\beta)(\gamma\delta)} - \\
\left(\beta J\right)^2 \left\{ \left[ \langle S^\alpha S^\beta S^\gamma S^\delta
\rangle\right]\z - \left[\langle  S^\alpha S^\beta \rangle\right]\z \,
\left[\langle S^\gamma S^\delta
\rangle \right]\z \right\} .
\label{A-Ising}
\end{multline}
where the average $\langle \cdots\rangle$ 
is over the spins for a fixed value of the Gaussian random field $z$
in Eq.~\eqref{Salpha_av}, and the average $[ \cdots]\z$ is over $z$ according to
Eq.~\eqref{fz}.

Because the theory is invariant under permutation of the replicas,
there are only three distinct
values for the matrix elements:
\begin{subequations}
\label{A}
\begin{align}
A_{(\alpha\beta),(\alpha\beta)} & = P , \\
A_{(\alpha\beta),(\alpha\gamma)} & = Q , \\
A_{(\alpha\beta),(\gamma\delta)} & = R , 
\end{align}
\end{subequations}
in which $\alpha, \beta, \gamma$ and $\delta$ are all different. Recall that
$(\alpha\beta)$ takes $n(n-1)/2$ distinct values,
i.e.~the pair $(\beta\alpha)$ is the same as the pair $(\alpha\beta)$.

\subsection{First eigenvalue and eigenvector}
\label{sec:l1}
If we go along \textit{any} row or column, the number of times, $P, Q$ and
$R$ appear is given by
\begin{subequations}
\begin{align}
n_P & = 1 , \\
n_Q & = 2(n-2) , \\
n_R & = \smfrac{1}{2}(n-2)(n-3) . 
\end{align}
\end{subequations}
Since the sum of all elements in any row or column is the same for each row
and column, it trivially follows that there is an eigenvector
\begin{equation}
\vec{e}_1 = (1, 1, \cdots, 1), 
\label{e1}
\end{equation}
with eigenvalue equal to the sum of all the elements along a row or column,
\begin{equation}
\lambda_1 = P + 2(n-2) Q + \smfrac{1}{2}(n-2)(n-3) R \, .
\label{lambda1}
\end{equation}
This eigenvalue has degeneracy 1.

\subsection{Second eigenvalue and eigenvectors}
\label{sec:lambda2}

We look for an eigenvector $\vec{e}_{2,\epsilon}$ with elements 
\begin{equation}
e_{2,\epsilon}^{\alpha\beta} = \left\{
\begin{array}{ll}
d & \text{(if $\alpha$ or $\beta = \epsilon$)}, \\
e & \text{(otherwise)}, \\
\end{array}
\right.
\label{e2}
\end{equation}
for some $\epsilon$.
The $\vec{e}_{2,\epsilon}$  must be orthogonal to $\vec{e}_1$ in Eq.~\eqref{e1},
which means
\begin{equation}
\sum_{(\alpha\beta)} e_{2,\epsilon}^{(\alpha\beta)} =  0,
\label{sume2}
\end{equation}
for each $\epsilon$,
and so
\begin{equation}
(n-2) e = - 2 d .
\label{de}
\end{equation}
Naively the there are $n$ \textit{independent} vectors
since there are $n$
choices for $\epsilon$. However, these are not all independent since it is
quite easy to show that
\begin{equation}
\sum_\epsilon \vec{e}_{2,\epsilon} = 0 .
\label{constraint}
\end{equation}
Hence there is one linear constraint among the $n$ vectors defined by
Eq.~\eqref{e2} and so the number of linearly independent such vectors is
$n-1$, i.e.~the degeneracy is $n-1$.
It is now straightforward to verify from Eqs.~\eqref{e2} and \eqref{de}, that 
\begin{equation}
A\, \vec{e}_{2,\epsilon} = \lambda_2\, \vec{e}_{2,\epsilon} ,
\end{equation}
where $\lambda_2$ is the eigenvalue, given by
\begin{equation}
\lambda_2 = P + (n-4) Q - (n-3) R .
\label{lambda2}
\end{equation}
Note that $\lambda_2 = \lambda_1$ for $n \to 0$.

\subsection{Third eigenvalue and eigenvectors}
\label{sec:l3}
We look for an eigenvector $\vec{e}_{3,(\eta\sigma)}$ with elements
\begin{equation}
e_{3,(\eta\sigma)}^{(\alpha\beta)} = \left\{
\begin{array}{ll}
f & \text{(if $(\alpha\beta)=(\eta\sigma)$)}, \\
g & \text{(if one of $(\alpha\beta)$ is equal to one of $(\eta\sigma)$)}, \\
h & \text{(if $(\alpha\beta) \ne (\eta\sigma)$)}, 
\end{array}
\right.
\label{e3}
\end{equation}
for some choice of $\eta$ and $\sigma$ (with $\sigma \ne \eta$).
The vectors in Eq.~\eqref{e3} must be orthogonal to $\vec{e}_1$ in Eq.~\eqref{e1},
and to the 
$\vec{e}_{2,\epsilon}$ in Eq.~\eqref{e2} so
\begin{equation}
f = (2 - n) g, \qquad g = \smfrac{1}{2}(3 - n) h .
\label{fghsol}
\end{equation}

One can show that summing over one of the indices labeling a vector,
gives zero, i.e.
\begin{equation}
\sum_\eta \vec{e}_{3,(\eta\sigma)} = 0.
\label{constraint3}
\end{equation}
Equation
\eqref{constraint3} gives $n$ constraints, one for each value of $\sigma$.
Hence the number of \textit{linearly independent} eigenvectors of the third
type is $n(n-1)/2$ (the number of values of the index $(\eta\sigma)$) less
$n$, the number of linear constraints. Hence the degeneracy is
$\smfrac{1}{2}n(n-3).$
One can also show that the sum over one of the replica component indices
vanishes for each vector\cite{bray:80b}, i.e.
\begin{equation}
\sum_\alpha e_{3,(\eta\sigma)}^{(\alpha\beta)}  = 0.
\label{sume3}
\end{equation}
(Recall that the subscript indices $(\eta\sigma)$ indicate a particular
vector, and the superscript indices $(\alpha\beta)$ denote a particular
element of that vector.)

It is now straightforward to show
that the vectors in Eq.~\eqref{e3} are indeed eigenvectors, i.e.
\begin{equation}
A\, \vec{e}_{3,(\eta\sigma)} = \lambda_3\, \vec{e}_{3,(\eta\sigma)} ,
\end{equation}
where $\lambda_3$ is the ``replicon'' eigenvalue,
\begin{equation}
\lambda_3 = P - 2 Q + R .
\label{lambda3}
\end{equation}

The total number of eigenvectors, of type 1, 2 or 3, found so far is
$1 + (n-1) + \smfrac{1}{2}n(n-3) = \smfrac{1}{2}n(n-1)$, which is the
dimension of the matrix. Hence we have found \textit{all} the eigenvalues and
eigenvectors.

\subsection{Matrix inverse}
\label{sec:inverse}
Consider the matrix $G$ which is the inverse of $A$, i.e.
\begin{equation}
A\, G = I \, 
\label{inverse}
\end{equation}
where $I$ is the identity matrix. We assume that $G$ has the same structure as
$A$ and define, see Eq.~\eqref{G123},
\begin{subequations}
\begin{align}
G_{(\alpha\beta),(\alpha\beta)} & = G_1 , \\
G_{(\alpha\beta),(\alpha\gamma)} & = G_2 , \\
G_{(\alpha\beta),(\gamma\delta)} & = G_3 , 
\end{align}
\end{subequations}
Evaluating the $(\alpha\beta),(\alpha\beta)$, the $(\alpha\beta),(\alpha\gamma) $,
and the $(\alpha\beta),(\gamma\delta)$ elements of Eq.~\eqref{inverse} gives
respectively
\begin{subequations}
\label{conds}
\begin{equation}
P\, G_1 + 2(n-2) Q\, G_2 + \smfrac{1}{2}(n-2)(n-3) R\, G_3 = 1 .
\label{cond1}
\end{equation}

\vspace{-0.8cm}
\begin{multline}
Q\, G_1 + \left[P + (n-2) Q + (n-3) R\right] G_2 + \\
\left[(n-3) Q + \smfrac{1}{2}(n-3)(n-4) R\right]  G_3 = 0 .
\label{cond2}
\end{multline}

\vspace{-0.8cm}
\begin{multline}
R\, G_1 + \left[4 Q + 2 (n-4) R \right] G_2 + \\
\hspace{0.4cm} \left[P + 2 (n-4) Q + \smfrac{1}{2}(n-4)(n-5) R\right]  G_3 = 0 .
\label{cond3}
\end{multline}
\end{subequations}
Taking $1 \times$\eqref{cond1}$\,-\,
2 \times$\eqref{cond2} $\,+\, 1 \times$\eqref{cond3} gives
\begin{equation}
(P - 2 Q + R) \, (G_1 - 2 G_2 + G_3) = 1 \, ,
\label{PQRG123}
\end{equation}
so the ``replicon propagator'' is given by
\begin{equation}
G_r \equiv G_1-2 G_2 +G_3 = {1 \over P - 2Q + R}\, .
\label{Gr}
\end{equation}
The spin glass susceptibility is determined from $G_r$ according to
Eq.~\eqref{chisg-Gr}. Note that Eqs.~\eqref{PQRG123} and
\eqref{chisg-Gr} determine $\chi_{SG}$ without needing to diagonalize the matrix
$A$. However, since the diagonalization has been done by AT it is instructive
to see that
$G_r$
\textit{is} the inverse of the replicon eigenvalue in Eq.~\eqref{lambda3}, see
also Appendix \ref{sec:diag-Ising}.

If we accept that $\lambda_3$ is an eigenvalue then Eq.~\eqref{Gr}
is obvious since the eigenvectors of $A$ and $G$ are the
same, and the corresponding eigenvalues are the inverses of each other.
Furthermore, since
the inverse matrix $G$
has the same structure as that of the original matrix $A$,
the expressions for the
eigenvalues of $A$ in terms of the parameters $P, Q$ and $R$, are the same as
the expressions for the eigenvalues of $G$ in terms of the corresponding
parameters $G_1, G_2$ and $G_3$.

\section{Fluctuation analysis for vector spin glasses}
\label{sec:diag-vector}

We now have additional indices for the spin components, and to avoid
confusion in notation, we will use Greek
letters $\alpha,\beta,\gamma,\delta,\epsilon$ for replicas and $\mu,\nu,\kappa,\sigma$
for spin indices. A row or column of the matrix will then involve 4
indices $(\alpha\beta), \mu\nu$, in which the order of the replica pair
$(\alpha\beta)$ is unimportant, i.e.~$(\beta\alpha)$
is the same as $(\alpha\beta)$,
but the order of the spin indices does matter because
$S_\alpha^\mu S_\beta^\nu$ is not the same as $S_\alpha^\nu S_\beta^\mu$.

Another new feature which appears when we deal with vector spins is the
appearance of terms with both replicas equal, $(\alpha\alpha)$.  These do not
appear for the Ising case because $\left(S_\alpha\right)^2$ is equal to 1, a
constant. However, $\left(S_\alpha^\mu\right)^2$ is not a constant for $m > 1$
and so we now need to include $(\alpha\alpha)$ terms
in the analysis.

Hence we shall need to find the eigenvalues and eigenvectors of a
matrix $Z$
of size $\smfrac{1}{2}n(n+1)\, m^2$ whose elements are given by
\begin{multline}
Z_{(\alpha\beta),(\gamma\delta)}^{\mu\nu,\kappa\sigma}
= \delta_{(\alpha\beta)(\gamma\delta)} \delta_{\mu\kappa} \delta_{\nu\sigma} - \\
\left(\beta J\right)^2 \left\{
\left[\langle S^\alpha_\mu S^\beta_\nu S^\gamma_\kappa S^\delta_\sigma
\rangle\right]\z - \left[\langle  S^\alpha_\mu S^\beta_\nu \rangle \right]\,
\left[\langle S^\gamma_\kappa S^\delta_\sigma
\rangle \right]\z
\right\}.
\label{Z-vector}
\end{multline}

Ignoring for now the spin indices
(which will be put back later) we consider the following matrix
of dimension $\smfrac{1}{2}n(n+1) \times \smfrac{1}{2}n(n+1)$,
\begin{equation}
Z =
\begin{pmatrix}
A & B \\
B^T & C \\
\end{pmatrix} ,
\label{Z}
\end{equation}
in which $A$ is the matrix of dimension $\smfrac{1}{2}n(n-1) \times\smfrac{1}{2} n(n-1)$
with rows and
columns labeled by two distinct replicas $(\alpha\beta)$ defined in
Eq.~\eqref{A},
$C$ is an $n \times
n$ matrix with rows and columns labeled by a single replica $(\alpha\alpha)$,
and $B$ is a matrix with $\smfrac{1}{2}n(n-1)$ rows and $n$ columns.

\subsection{Decomposing into subspaces}
\label{sec:subspaces}

We now discuss each of these matrices in turn.
\begin{itemize}
\item
The elements of $A$ are given by Eq.~\eqref{A}.
\item
The elements of $B$ are
\begin{subequations}
\label{B}
\begin{align}
B_{(\alpha\beta),(\alpha\alpha)} & = S , \\
B_{(\alpha\beta),(\gamma\gamma)} & = T , 
\end{align}
\label{BST}
\end{subequations}
in which $\alpha,\beta$ and $\gamma$ are all different.
\item
The elements of $C$ are
\begin{subequations}
\label{C}
\begin{align}
C_{(\alpha\alpha),(\alpha\alpha)} & = U , \\
C_{(\alpha\alpha),(\beta\beta)} & = V , 
\end{align}
\end{subequations}
in which $\alpha$ and $\beta$ are different.
\end{itemize}

Now we add the Cartesian spin indices.
The result is that each element of the
matrix $Z$ in Eq.~\eqref{Z}
becomes an $m^2 \times m^2$ matrix with rows and columns labeled
by a pair of spin component indices $\mu$ and $\nu$, each of which runs over
values from 1 to $m$.

A simplification is that the only non-zero elements
are those where each Cartesian spin component occurs an even
number of times (combining the row and column indices). Hence each $m^2
\times m^2$ matrix
breaks up into different blocks. There is one $m \times m$ block,
$ (\mu\mu, \nu\nu) $
where $ \mu = 1, \cdots, n,\ \nu = 1, \cdots m, $
and $m(m-1)/2$ blocks of size 2,
$ (\mu\nu, \mu\nu)$ and $(\mu\nu, \nu\mu) $
where $\mu $ and $\nu \ (\ne \mu)$
are fixed.

Consider, for example, one of the elements in $A$ with value $P$, see
Eq.~\eqref{A}. This is now expanded into an $m^2 \times m^2$ matrix which
is block diagonalized into
\begin{itemize}
\item
(i) one $m \times m$ matrix, with rows and columns labeled by
$\mu\mu\, (\mu =
1, \cdots, m)$,
\begin{equation}
\begin {pmatrix}
P_L & P_T \cdots P_T \\
P_T & P_L \cdots P_T \\
\vdots & \vdots \ddots \vdots \\
P_T & P_T \cdots P_L \\
\end {pmatrix} ,
\label{PLPT}
\end{equation}
where the diagonal elements (to which we give the subscript $L$) are different
from the off-diagonal elements (to which we give the subscript $T$), and
\item
(ii)
$m(m-1)/2$ identical matrices of size $2 \times 2$, with rows and columns
labeled by $\mu\nu$ and $\nu\mu$ (for fixed $\mu$ and $\nu$ with $\mu\ne\nu$),
\begin{equation}
\begin {pmatrix}
P_1 & P_2 \\
P_2 & P_1  \\
\end {pmatrix} ,
\label{P1P2}
\end{equation}
in which we give the subscript ``1'' to the (equal) diagonal elements and the
subscript ``2'' to the off-diagonal elements.
\end{itemize}
The eigenvalues of \eqref{PLPT} are
\begin{subequations}
\label{PSA}
\begin{align}
P_S & = P_L + (m-1) P_T, \quad\,(\text{degeneracy\ } 1), \\
P_A & = P_L - P_T,  \qquad\qquad\ \ (\text{degeneracy\ } m-1) ,
\end{align}
\end{subequations}
and those of \eqref{P1P2} are
\begin{subequations}
\label{l+-}
\begin{align}
P_+ & = P_1 + P_2,
\label{l+}\\
P_- & = P_1 - P_2 ,
\label{l-}
\end{align}
\end{subequations}
each of degeneracy 1.

The $R,S,T,U$ and $V$
elements of the replica matrix,
in
Eqs.~\eqref{A}, \eqref{B} and \eqref{C},
expand out into the same block structure in
spin-component space.

However, we shall now show that
things are somewhat different for the $Q$
%
elements,
which have replica structure $(\alpha\beta),(\alpha\gamma)$,
i.e.~\textit{one}
of the replicas is repeated. The order of the replica indices in a pair
does not matter
but, to keep track of which spin index goes with which replica index,
we should adopt some convention, e.g.~put the lower replica 
index first. Consider then
a situation with $\beta < \gamma$ and different values of $\alpha$. The $Q$
element involving these three replicas
would then be labeled differently depending on the value of $\alpha$
relative to $\beta$ and $\gamma$ as follows:
\begin{subequations}
\begin{align}
(\alpha\beta), (\alpha\gamma) \quad (\alpha < \beta < \gamma), \\
(\beta\alpha), (\alpha\gamma) \quad (\beta < \alpha < \gamma), \\
(\beta\alpha), (\gamma\alpha) \quad (\beta < \gamma< \alpha ), 
\end{align}
\end{subequations}
Hence the $2 \times 2$ matrix $Q$ has the form 
\begin{equation}
\begin{array}{l|l l | l}
       & \mu\nu & \nu\mu \\
\hline
\mu\nu & Q_1    & Q_2    \\
\nu\mu & Q_2    & Q_1    \\
\end {array} 
\label{Q12}
\end{equation}
for $\alpha < \beta < \gamma$ and 
$\beta < \gamma< \alpha $, 
while it is 
\begin{equation}
\begin{array}{l|l l | l}
       & \mu\nu & \nu\mu \\
\hline
\mu\nu & Q_2    & Q_1    \\
\nu\mu & Q_1    & Q_2    \\
\end {array} 
\label{Q21}
\end{equation}
for $\beta < \alpha < \gamma$ , 
i.e.~$Q_1$ and $Q_2$ are interchanged in the latter case.
This does not affect $Q_+ \equiv Q_1 + Q_2$ but it changes the sign of $Q_-$
when the repeated replica index  ($\alpha$ here) lies in between the other two
($\beta$ and $\gamma$ here).

Our goal is to diagonalize the matrix $Z$ given by Eq.~\eqref{Z}, in
which $A, B$ and $C$, are matrices in replica space given by Eqs.~\eqref{A},
\eqref{B} and \eqref{C}, and each element in these matrices is itself an $m^2
\times m^2$ replica in spin-component space which block diagonalizes
as discussed above.
Symbolically we want to find the eigenvalues
$\lambda$ and eigenvectors $(\vec{e}, \vec{f})$ of
\begin{equation}
\begin{pmatrix}
A & B \\
B^T & C \\
\end{pmatrix}
\begin{pmatrix}
\vec{e} \\
\vec{f} \\
\end{pmatrix}
 = \lambda
\begin{pmatrix}
\vec{e} \\
\vec{f} \\
\end{pmatrix} ,
\label{evaleq}
\end{equation}
where the vector $\vec{e}$ is of dimension $\smfrac{1}{2}n(n-1) m^2$
and $\vec{f}$ is of dimension $n m^2$.

Because the block structure in spin-component space is the same for all
elements of $Z$ in
Eq.~\eqref{Z} (except for the some aspects of the ``$-$''
sector),,
we can diagonalize separately the spin-component and
replica sectors.
Hence the eigenvalue equation, Eq.~\eqref{evaleq},
breaks up into 4 simpler sets equations, one set for
each distinct spin-component sector:
\begin{itemize}
\item
1 set of equations of the type
\begin{equation}
\begin{pmatrix}
A_S & B_S \\
B_S^T & C_S \\
\end{pmatrix}
\begin{pmatrix}
\vec{e}_S \\
\vec{f}_S \\
\end{pmatrix}
 = \lambda_S
\begin{pmatrix}
\vec{e}_S \\
\vec{f}_S \\
\end{pmatrix} ,
\label{evaleqS}
\end{equation}
\item
$m-1$ identical sets of equations of the type
\begin{equation}
\begin{pmatrix}
A_A & B_A \\
B_A^T & C_A \\
\end{pmatrix}
\begin{pmatrix}
\vec{e}_A \\
\vec{f}_A \\
\end{pmatrix}
 = \lambda_A
\begin{pmatrix}
\vec{e}_A \\
\vec{f}_A \\
\end{pmatrix} ,
\label{evaleqA}
\end{equation}
\item
$m(m-1)/2$ identical sets of equations of the type
\begin{equation}
\begin{pmatrix}
A_+ & B_+ \\
B_+^T & C_+ \\
\end{pmatrix}
\begin{pmatrix}
\vec{e}_+ \\
\vec{f}_+ \\
\end{pmatrix}
 = \lambda_+
\begin{pmatrix}
\vec{e}_+ \\
\vec{f}_+ \\
\end{pmatrix} ,
\label{evaleq+}
\end{equation}
\item
and $m(m-1)/2$ identical sets of equations of the type
\begin{equation}
\begin{pmatrix}
A_- & B_- \\
B_-^T & C_- \\
\end{pmatrix}
\begin{pmatrix}
\vec{e}_- \\
\vec{f}_- \\
\end{pmatrix}
 = \lambda_-
\begin{pmatrix}
\vec{e}_- \\
\vec{f}_- \\
\end{pmatrix} .
\label{evaleq-}
\end{equation}
\end{itemize}
The matrices in Eqs.~\eqref{evaleqS}--\eqref{evaleq-} are
of dimension $n(n+1)/2$, while the vectors $\vec{e}$ are of length $n(n-1)/2$
and the vectors $\vec{f}$ are of length $n$.


Each of the sets of equations,~\eqref{evaleqS}--\eqref{evaleq-}
has the same structure,
which is a little more complicated than diagonalizing the
matrix $A$, described in the first part of this report, because the
off-diagonal piece $B$ couples the elements of $A$ to
the $n\times n$ block $C$, in which each row or
column index has two equal replicas.
However, we shall see that the square blocks $A$ and $C$ decouple in two
cases: (i) the ``$-$'' sector, and
(ii) the replicon eigenvectors in the $S, A$, and ``$+$'' sectors.

We shall first discuss the $S, A$ and ``$+$'' spin-component
sectors together, and then do the
``$-$'' sector which has to be treated separately.


\subsection{The $\boldsymbol{S, A}$ and ``$\boldsymbol{+}$'' Spin-Component Sectors}
The matrices for these sectors are all the same provided one replaces the
elements of the 
replica matrix $Z$ in
Eq.~\eqref{Z} by the appropriate eigenvalue of the spin-component sector,
see Eqs.~\eqref{PSA}--\eqref{l+-} for the case of $P$.

We first discuss the replicon subspace.
\subsubsection{Replicon Modes}

Let us see if the replicon eigenvector,
computed for the Ising case in
Sec.~\ref{sec:l3}, satisfies Eq.~\eqref{Z} with $\vec{f} = 0$, i.e.
\begin{equation}
\begin{pmatrix}
A & B \\
B^T & C \\
\end{pmatrix}
\begin{pmatrix}
\vec{e}_3 \\
0 \\
\end{pmatrix}
 = \lambda_3
\begin{pmatrix}
\vec{e}_3 \\
0 \\
\end{pmatrix} ,
\label{evaleql3}
\end{equation}
which requires
\begin{equation}
\sum_{(\alpha\beta)} B_{(\alpha\beta), (\gamma\gamma)} e_3^{(\alpha\beta)} =
0,
\label{decouple}
\end{equation}
for each $\gamma$.
From Eq.~\eqref{B} we have 
\begin{equation}
\sum_{(\alpha\beta)} B_{(\alpha\beta), (\gamma\gamma)} e_3^{(\alpha\beta)} =
2 S \sum_\beta e_3^{(\gamma\beta)} + T \sum_{\alpha\ne\gamma,\beta\ne\gamma}
e_3^{(\alpha\beta)}.
\end{equation}
The first term vanishes because of Eq.~\eqref{sume3}. Again using
Eq.~\eqref{sume3}, the sum in the second term can be written as
$-\sum_{\beta\ne\gamma} e_3^{(\gamma\beta)}$, which again vanishes by
Eq.~\eqref{sume3}. Hence Eq.~\eqref{decouple} is satisfied.

As a result, we don't need to do any more work to get the eigenvalues
in the replicon sector for the vector spin glass. We just use
Eq.~\eqref{lambda3} for
each of the $S, A$ and ``$+$''
spin-component sectors in Eqs.~\eqref{evaleqS}--\eqref{evaleq+}, i.e.
\begin{subequations}
\label{l3vec}
\begin{align}
\lambda_{3S} &= P_S - 2 Q_S + R_S \nonumber  \\
&= [P_L +(m-1)P_T] -2[Q_L +(m-1)Q_T] +
\nonumber \\
&\hspace{2cm} [R_L +(m-1)R_T] ,
\label{l3vecS}
\\
\lambda_{3A} &= P_A - 2 Q_A + R_A  \nonumber \\
&= (P_L - P_T) -2(Q_L - Q_T),
+ (R_L - R_T) \label{l3A} \\ 
\lambda_{3+} &= P_+ - 2 Q_+ + R_+  \nonumber \\
&= (P_1 + P_2) -2(Q_1 + Q_2)
+ (R_1 + R_2) .
\label{l3+}
\end{align}
\end{subequations}
As discussed above, the spin-component degeneracies of the $S, A,$ and ''$+$''
subspaces are $1, (m-1),$ and $\smfrac{1}{2}m(m-1)$
respectively.
To get the overall degeneracies of the
eigenvalues in Eq.~\eqref{l3vec} one has to multiply these factors by the
degeneracy in replica space, $\smfrac{1}{2}n(n-3)$.


\subsubsection{``$\lambda_1$'' Modes}

Referring to Eq.~\eqref{evaleqS} we look for a solution where all the elements of
$\vec{e}_{S1}$ are equal to $a$, say and all the elements of $\vec{f}_{S1}$ are equal
to $b$. This gives the coupled equations
\begin{equation}
\begin{pmatrix}
\alpha_{1S} & \beta_{1S} \\
\smfrac{1}{2}(n-1) \beta_{1S} & \gamma_{1S}
\end{pmatrix}
\begin{pmatrix}
a \\ b
\end{pmatrix}
=\lambda_{1S} 
\begin{pmatrix}
a \\ b 
\end{pmatrix}
\end{equation}
where
\begin{subequations}
\label{abc1S}
\begin{align}
\alpha_{1S} &= [P_L + (m-1)P_T] +2(n-2) [Q_L + (m-1)Q_T] + \nonumber \\
& \hspace{1cm} \smfrac{1}{2}(n-2)(n-3)
[R_L + (m-1)R_T] , \\
\beta_{1S} &= 2 [S_L +(m-1)S_T] + (n-2) [T_L +(m-1)T_T] , \\
\gamma_{1S} &= [U_L + (m-1) U_T] + (n-1)[V_L + (m-1) V_T] .
\end{align}
\end{subequations}
The eigenvalues are given by the solutions of the resulting quadratic equation
\begin{subequations}
\begin{align}
\lambda_{1a,S} &= {1\over2}\, \left[\alpha_{1S} + \gamma_{1S} +
\sqrt{(\alpha_{1S} - \gamma_{1S})^2 + 2(n-1)\beta_{1S}^2} \right] , \\
\lambda_{1b,S} &= {1\over2}\, \left[\alpha_{1S} + \gamma_{1S} -
\sqrt{(\alpha_{1S} - \gamma_{1S})^2 + 2(n-1)\beta_{1S}^2} \right] . 
\end{align}
\end{subequations}

This calculation simply repeats for the $A$ and ``$+$'' sectors with the
appropriate substitutions for $\alpha, \beta$ and $\gamma$. The 
spin-component degeneracies for the $S$, $A$ and ``$+$'' sectors
are 1, $m-1$ and $m(m-1)/2$ respectively. These have to be multiplied by the
degeneracy from the replica sector, which is just 1 in this case.
%

\subsubsection{``$\lambda_2$'' Modes}

We follow the procedure of Sec.~\ref{sec:lambda2}, by looking for an
eigenvector in which one replica, $\epsilon$ say, is distinct from the others.
Referring to Eq.~\eqref{evaleqA}, we set $\vec{e}_{S2}^{\alpha\beta}$ equal to $d$ if
$\alpha$ or $\beta$ are equal to $\epsilon$, and equal to $e$ otherwise.
Orthogonality to the $\lambda_1$ eigenvector requires $(n-2) e = -2d$, see
Eq.~\eqref{de}. Similarly we set $\vec{f}_{S2}^\alpha$ equal to $f$ if $\alpha =
\epsilon$ and equal to $g$ otherwise. Orthogonality to the $\lambda_1$
eigenvector requires $(n-1) g = -f$.

Substituting into Eq.~\eqref{evaleqA} then gives the coupled equations
\begin{equation}
\begin{pmatrix}
\alpha_{2S} & \beta_{2S} \\
(n-2) \beta_{2S} & \gamma_{2S}
\end{pmatrix}
\begin{pmatrix}
d \\ g
\end{pmatrix}
=\lambda_{1S} 
\begin{pmatrix}
d \\ g 
\end{pmatrix}
\end{equation}
where
\begin{subequations}
\label{abc2S}
\begin{align}
\alpha_{2S} &= [P_L + (m-1)P_T] +(n-4) [Q_L + (m-1)Q_T] \nonumber \\
& \hspace{1cm} -
(n-3) [R_L + (m-1)R_T] , \\
\beta_{2S} &= [S_L +(m-1)S_T] - [T_L +(m-1)T_T] , \\
\gamma_{2S} &= [U_L + (m-1) U_T] - [V_L + (m-1) V_T] .
\end{align}
\end{subequations}

The eigenvalues are given by the solutions of the resulting quadratic equation
\begin{subequations}
\begin{align}
\lambda_{2a,S} &= {1\over2}\, \left[\alpha_{2S} + \gamma_{2S} +
\sqrt{(\alpha_{2S} - \gamma_{2S})^2 + 4(n-2)\beta_{2S}^2} \right] , \\
\lambda_{2b,S} &= {1\over2}\, \left[\alpha_{2S} + \gamma_{2S} -
\sqrt{(\alpha_{2S} - \gamma_{2S})^2 + 4(n-2)\beta_{2S}^2} \right] . 
\end{align}
\end{subequations}
Analogous
results are obtained for the $A$ and ``$+$'' subspaces. The spin-component
degeneracies for the $S$, $A$ and ``$+$'' subspaces are 1,  $m-1$ and $m(m-1)/2$ respectively.
These have to be multiplied by the degeneracy from the replica sector which is
$n-1$.

%
%

\subsection{The ``$\boldsymbol{-}$'' Spin-Component Sector}
\label{sec:-}

The spin-component degeneracy for these eigenvalues is
$\smfrac{1}{2}m(m-1)$.

The blocks with $(\alpha\beta)$
($\beta\ne\alpha$) and
$(\alpha\alpha)$ decouple. The reason is that~\cite{almeida:80}
$S_1 = S_2$ and $T_1 = T_2$ which follows from the symmetry properties of the
expression for the matrix elements in Eq.~\eqref{Z-vector} and the definitions in
Eqs.~\eqref{BST} and \eqref{P1P2}.
It then follows from Eq.~\eqref{l+-} that $S_- = T_- = 0$ and
so, from Eq.~\eqref{B}, the matrix $B_-$ vanishes.


\subsubsection{The $(\alpha\alpha)$ Subspace}

We can easily obtain the ``$-$'' eigenvalue which lies entirely within the
$(\alpha\alpha)$ subspace, since~\cite{almeida:80}
$V_1 = V_2$,  so $V_- = 0$
and hence $C_-$ is a diagonal $n \times n$ matrix with the constant
value $U_- = U_1 - U_2$ on
the diagonal.  Hence there is an eigenvalue
\begin{equation}
\lambda_{4-} = U_- = U_1 - U_2 \, ,
\end{equation}
with total degeneracy $\smfrac{1}{2}n\, m(m-1)$.
It has no analogue in the other spin-component sectors.

\subsubsection{The $(\alpha\beta)$ Subspace,
$(\alpha  \ne\beta)$}

Now we consider the ``$-$'' eigenvalues which lie entirely within the
$(\alpha\beta)$ ($\beta\ne\alpha$) subspace of dimension $n(n-1)/2$. We find
that there are two distinct eigenvalues (for general $n$).

We ask if there are eigenvalues analogous to $\lambda_1, \lambda_2, \lambda_3$
that we found in Appendix \ref{sec:diag-Ising} for the Ising case.
The eigenvector corresponding to
$\lambda_1$ for the Ising case has all components equal, since the sum along all
rows and columns of the matrix is the same. However, this is not the case here
because the number of times $-Q$ occurs is therefore
different for different rows or columns.
Hence there is no eigenvalue analogous to $\lambda_1$.

Each of the $n-1$ eigenvectors for $\lambda_2$ for the Ising case singled out a
particular replica.
For example, picking out out replica $\epsilon$, the coefficients of
(i) $(\epsilon\alpha)$, and (ii) $(\alpha\beta)$ in which
neither $\alpha$ nor $\beta$ equal to $\epsilon$,
would be different, see Eq.~\eqref{de}.  There is a
similar eigenvector here in which the type (ii) components vanish and the type
(i) components are no longer all equal but have value $-1$ for $\alpha$ less
than the special replica ($\epsilon$ in our example),
and $+1$ for $\alpha$ greater than the special replica.
By inspection this has eigenvalue 
\begin{equation}
\lambda_{2-} = P + (n-2) Q \, .
\end{equation}
There are $n$ ways to pick the special replica, but the resulting $n$
eigenvectors sum to zero
since, in the sum, each element appears twice, once with a plus sign and once
with a minus sign.
In other words, there is one
linear relation between the eigenvectors,
so the replica degeneracy of $\lambda_2$ is
actually $n-1$ rather than $n$.


For the Ising case, each of the eigenvectors for $\lambda_3$ picked out 2
replicas. For example, picking out replicas $\alpha$ and $\beta$, the coefficients of
(i) $(\alpha\beta)$, (ii) $(\alpha\gamma)$ and $(\beta\gamma)$, and
(iii) $(\gamma\delta)$ (all replicas with a different label assumed different)
would be different, see
Eq.~\eqref{fghsol}. There are eigenvectors like this here, in which type (iii)
components are zero, type (ii) components  $(\alpha\gamma)$ have value
$1$ if $\gamma< \alpha < \beta$
and $-1$ otherwise, 
and the type (i) component
has value $n-2$. For example, for $n=4$ there are 6 such vectors (not all
independent, see below) which are
\begin{align}
\vec{e}_{(12)} = & \quad ( 2, -1, -1,  1,  1,  0) , \nonumber \\
\vec{e}_{(13)} = & \quad (-1,  2, -1, -1,  0,  1) , \nonumber \\
\vec{e}_{(14)} = & \quad (-1, -1,  2,  0, -1, -1) , \nonumber \\
\vec{e}_{(23)} = & \quad ( 1, -1,  0,  2, -1,  1) , \nonumber \\
\vec{e}_{(24)} = & \quad ( 1,  0, -1, -1,  2, -1) , \nonumber \\
\vec{e}_{(34)} = & \quad ( 0,  1, -1,  1, -1,  2) . 
\end{align}
By inspection these have eigenvalue
\begin{equation}
\lambda_{3-} = P - 2 Q \, .
\end{equation}
One can also see from the above vectors, which are for $n=4$, that
$\vec{e}_{(14)}$ can be expressed as a linear combination of the
$\vec{e}_{(1\alpha)}$ for $\alpha < 4$, and similarly $\vec{e}_{(24)}$ can be expressed as a
linear combination of the
$\vec{e}_{(2\alpha)}$ for $\alpha < 4$, and the same for $\vec{e}_{(34)}$.
Hence the last replica can be eliminated, so the number of
\textit{linearly independent} vectors, which is the replica degeneracy of 
$\lambda_{3-}$, is $\smfrac{1}{2}(n-1)(n-2)$.

Hence, including both $\lambda_{2-}$ and $\lambda_{3-}$, we have found all
$\smfrac{1}{2}n(n-1)$
eigenvalues and eigenvectors in the replica sector.

\subsection{Summary of Eigenvalues}

\begin{table*}
\begin{tabular}{||l|c|c|c||}
\hline\hline
\multicolumn{2}{|c|}{
Eigenvalue } & replica degeneracy & spin-component degeneracy \\
\hline\hline
$\lambda_{1a, S} $&
${1\over2}\, \left[\alpha_{1S} + \gamma_{1S} +
\sqrt{(\alpha_{1S} - \gamma_{1S})^2 + 2(n-1)\beta_{1S}^2} \right] $& 1              & 1 \\
$\lambda_{1b, S} $&
${1\over2}\, \left[\alpha_{1S} + \gamma_{1S} -
\sqrt{(\alpha_{1S} - \gamma_{1S})^2 + 2(n-1)\beta_{1S}^2} \right]$& 1               & 1 \\
$\lambda_{2a, S} $&
${1\over2}\, \left[\alpha_{2S} + \gamma_{2S} +
\sqrt{(\alpha_{2S} - \gamma_{2S})^2 + 4(n-2)\beta_{2S}^2} \right] $& $n-1$             & 1 \\
$\lambda_{2b, S} $&
${1\over2}\, \left[\alpha_{2S} + \gamma_{2S} -
\sqrt{(\alpha_{2S} - \gamma_{2S})^2 + 4(n-2)\beta_{2S}^2} \right] $& $n-1$                   & 1 \\
$\lambda_{3S}  $&$[P_L +(m-1)P_T] -2[Q_L +(m-1)Q_T] + [R_L +(m-1)R_T]$&$ \smfrac{1}{2}n(n-3) $  & 1 \\
\hline
$\lambda_{1a, A} $&
$ {1\over2}\, \left[\alpha_{1A} + \gamma_{1A} +
 \sqrt{(\alpha_{1A} - \gamma_{1A})^2 + 2(n-1)\beta_{1A}^2} \right]$& 1            &$ m-1$ \\
$\lambda_{1b, A} $&
${1\over2}\, \left[\alpha_{1A} + \gamma_{1A} -
\sqrt{(\alpha_{1A} - \gamma_{1A})^2 + 2(n-1)\beta_{1A}^2} \right]$& 1             &$ m-1$ \\
$\lambda_{2a, A} $&
${1\over2}\, \left[\alpha_{2A} + \gamma_{2A} +
\sqrt{(\alpha_{2A} - \gamma_{2A})^2 + 4(n-2)\beta_{2A}^2} \right]
$& $n-1$                   &$ m-1$ \\
$\lambda_{2b, A} $&
${1\over2}\, \left[\alpha_{2A} + \gamma_{2A} -
\sqrt{(\alpha_{2A} - \gamma_{2A})^2 + 4(n-2)\beta_{2A}^2} \right]
$& $n-1 $                  & $m-1$ \\
$\lambda_{3A}  $&$(P_L - P_T) -2(Q_L - Q_T) + (R_L - R_T)$& $\smfrac{1}{2}n(n-3) $  &$ m-1$ \\
\hline
$\lambda_{1a, +} $&
$ {1\over2}\, \left[\alpha_{1+} + \gamma_{1+} +
 \sqrt{(\alpha_{1+} - \gamma_{1+})^2 + 2(n-1)\beta_{1+}^2} \right] $& 1 & $\smfrac{1}{2}m(m-1)$ \\
$\lambda_{1b, +} $&
${1\over2}\, \left[\alpha_{1+} + \gamma_{1+} -
\sqrt{(\alpha_{1+} - \gamma_{1+})^2 + 2(n-1)\beta_{1+}^2} \right] $& 1 &$ \smfrac{1}{2}m(m-1)$ \\
$\lambda_{2a, +} $&
${1\over2}\, \left[\alpha_{2+} + \gamma_{2+} +
\sqrt{(\alpha_{2+} - \gamma_{2+})^2 + 4(n-2)\beta_{2+}^2} \right]
$& $n-1 $                  & $\smfrac{1}{2}m(m-1)$ \\
$\lambda_{2b, +} $&
${1\over2}\, \left[\alpha_{2-} + \gamma_{2-} +
\sqrt{(\alpha_{2-} - \gamma_{2-})^2 + 4(n-2)\beta_{2-}^2} \right]
$& $n-1 $                  & $\smfrac{1}{2}m(m-1)$ \\
$\lambda_{3+}  $&$(P_1 + P_2) -2(Q_1 + Q_2) + (R_1 + R_2)$& $\smfrac{1}{2}n(n-3) $  &$ \smfrac{1}{2}m(m-1)$ \\
\hline
$\lambda_{2-}  $&$(P_1 - P_2) + (n-2)(Q_1 - Q_2) $& $n-1$ & $\smfrac{1}{2}m(m-1)$ \\
$\lambda_{3-}  $&$(P_1 - P_2) -2(Q_1 - Q_2) $& $\smfrac{1}{2}(n-1)(n-2)$ &$ \smfrac{1}{2}m(m-1)$ \\
$\lambda_{4-}  $&$ U_1 - U_2                $& $n$                   &$ \smfrac{1}{2}m(m-1)$ \\
\hline\hline
\end{tabular}
\caption{Eigenvalues and degeneracies for the $m$-component model. The total
degeneracy for each eigenvalue is the product of the replica degeneracy and the
spin-component degeneracy. It is easy to see that the total degeneracy is
$\smfrac{1}{2}n(n+1) \times m^2$, as required.
\label{tab:vector}
}
\end{table*}

The eigenvalues, along with their degeneracies,
are summarized in Table~\ref{tab:vector},
in which the $\alpha$'s, $\beta$'s and $\gamma$'s are defined by
\begin{subequations}
\begin{align}
\alpha_{1S} &= [P_L + (m-1)P_T] +2(n-2) [Q_L + (m-1)Q_T] \nonumber \\
& \hspace{2cm} +
\smfrac{1}{2}(n-2)(n-3)
[R_L + (m-1)R_T] , \\
\beta_{1S} &= 2 [S_L +(m-1)S_T] + (n-2) [T_L +(m-1)T_T] , \\
\gamma_{1S} &= [U_L + (m-1) U_T] + (n-1)[V_L + (m-1) V_T] , 
\end{align}
\begin{align}
\alpha_{1A} &= (P_L - P_T) +2(n-2) (Q_L - Q_T) \nonumber \\
& \hspace{2cm} + \smfrac{1}{2}(n-2)(n-3)
(R_L - R_T) , \\
\beta_{1A} &= 2 (S_L - S_T) + (n-2) (T_L - T_T) , \\
\gamma_{1A} &= (U_L - U_T) + (n-1)(V_L - V_T) , 
\end{align}
\begin{align}
\alpha_{1+} &= (P_1 + P_2) +2(n-2) (Q_1 + Q_2) \nonumber \\
& \hspace{2cm} + \smfrac{1}{2}(n-2)(n-3)
(R_1 + R_2) , \\
\beta_{1+} &= 2 (S_1 + S_2) + (n-2) (T_1 + T_2) , \\
\gamma_{1+} &= (U_1 + U_2) + (n-1)(V_1 + V_2) ,
\end{align}
\begin{align}
\alpha_{2S} &= [P_L + (m-1)P_T] +(n-4) [Q_L + (m-1)Q_T] \nonumber \\
& \hspace{2.2cm} -
(n-3) [R_L + (m-1)R_T] , \\
\beta_{2S} &= [S_L +(m-1)S_T] - [T_L +(m-1)T_T] , \\
\gamma_{2S} &= [U_L + (m-1) U_T] - [V_L + (m-1) V_T] , 
\end{align}
\begin{align}
\alpha_{2A} &= (P_L - P_T) +(n-4) (Q_L - Q_T) \nonumber \\
& \hspace{3cm} - (n-3) (R_L - R_T) , \\
\beta_{2A} &= (S_L - S_T) - (T_L - T_T) , \\
\gamma_{2A} &= (U_L - U_T) - (V_L - V_T) , 
\end{align}
\begin{align}
\alpha_{2+} &= (P_1 + P_2) +(n-4) (Q_1 + Q_2) \nonumber \\
& \hspace{3cm} - (n-3) (R_1 + R_2) , \\
\beta_{2+} &= (S_1 + S_2) - (T_1 + T_2) , \\
\gamma_{2+} &= (U_1 + U_2) - (V_1 + V_2) .
\end{align}
\end{subequations}
By symmetry\cite{almeida:80},
$R_1 = R_2, S_1 = S_2, T_1 = T_2, V_1 = V_2$.

There are 18 distinct eigenvalues for arbitrary $n$. In the limit $n\to0$,
$\alpha_{1S}=\alpha_{2S}$,
$\beta_{1S}=\beta_{2S}$,
$\gamma_{1S}=\gamma_{2S}$,
$\alpha_{1A}=\alpha_{2A}$,
$\beta_{1A}=\beta_{2A}$,
$\gamma_{1A}=\gamma_{2A}$,
$\alpha_{1+}=\alpha_{2+}$,
$\beta_{1+}=\beta_{2+}$,
$\gamma_{1+}=\gamma_{2+}$, so $\lambda_{1a,S} = \lambda_{2a,S}$, etc., and
also $\lambda_{2-} = \lambda_{3-}$.  Hence there are only 11 distinct
eigenvalues in the $n \to 0$ limit.

Most of the results in Table~\ref{tab:vector} agree with those in
de Almeida's
thesis\cite{almeida:80}. However, there are some differences, the most notable
of which is that the eigenvalue $\lambda_{3S}$, which gives the divergence
of the non-linear susceptibility according to Eq.~\eqref{chisg_l3},
does not appear in
Ref.~\onlinecite{almeida:80}. However, we are confident that this eigenvalue
is correct and that its change of sign gives the AT instability. 
We note, for example, that the combination of propagators
on the LHS of Eq.~\eqref{G-vector} corresponds precisely to that in Eq.~(3.5) of
Ref.~\onlinecite{bray:79b}.

\vspace{0.5cm}
In the Ising ($m=1$) limit only the ``$S$'' eigenvalues survive (the degeneracy
of the rest is zero), and we present these results in Table~\ref{tab:Ising}. 
One has\cite{almeida:80}
$S_L = T_L = 0$, so
$\beta_{1S} = 0$, and $U_L = 1, V_L = 0$, so $\gamma_{1S} = 1$.
Hence the first four eigenvalues are
$\alpha_{1S}, 1, \alpha_{2S}$ and 1.  The two eigenvalues that are equal to 1
involve fluctuations of the $q_{\alpha\alpha}$ which couple to
$\left(S_\alpha\right)^2$, a constant, (so there is no actual coupling to the
spins). Hence these
eigenvalues are trivial.
The remaining three eigenvalues are just those of the original
AT paper~\cite{almeidaAT:78}, see Table~\ref{tab:Ising}.

\begin{table}[tbh]
\begin{tabular}{||l|c|c|}
\hline\hline
\multicolumn{2}{|c|}{
Eigenvalue} & degeneracy \\
\hline\hline
$\lambda_{1a}$  & $ P_L +2(n-2)Q_L + \smfrac{1}{2}(n-2)(n-3) R_L $  & 1   \\
$\lambda_{1b}$  & $ 1                                  $   & 1   \\
$\lambda_{2a}$  & $ P_L + (n-4) Q_L - (n-3) R_L        $   & $n-1$ \\
$\lambda_{2b}$  & $ 1                                  $   & $n-1$ \\
$\lambda_3   $  & $ P_L - 2 Q_L + R_L                  $   & $\smfrac{1}{2}n(n-3)$\\
\hline\hline
\end{tabular}
\caption{Eigenvalues and degeneracies for the Ising case, $m=1$.
\label{tab:Ising}}
\end{table}


\subsection{Matrix inverse for vector case}
\label{sec:inverse-vector}
As for the Ising case, we assume that 
$G$, the matrix inverse of $Z$,
has the same structure as $Z$ itself. In particular, we define
\begin{subequations}
\begin{align}
G_{(\alpha\beta),(\alpha\beta)}^{\mu\mu,\mu\mu}  & = G_{1L}, \ \ 
G_{(\alpha\beta),(\alpha\beta)}^{\mu\mu,\nu\nu}   = G_{1T} \ (\mu\ne \nu), \\
G_{(\alpha\beta),(\alpha\gamma)}^{\mu\mu,\mu\mu} & = G_{2L},  \ \ 
G_{(\alpha\beta),(\alpha\gamma)}^{\mu\mu,\nu\nu}  = G_{2T} \ (\mu \ne \nu), \\
G_{(\alpha\beta),(\gamma\delta)}^{\mu\mu,\mu\mu} & = G_{3L}, \ \ 
G_{(\alpha\beta),(\gamma\delta)}^{\mu\mu,\nu\nu}  = G_{3T} \ (\mu\ne \nu),  \\
G_{(\alpha\beta),(\alpha\alpha)}^{\mu\mu,\mu\mu} & = G_{4L}, \ \ 
G_{(\alpha\beta),(\alpha\alpha)}^{\mu\mu,\nu\nu}  = G_{4T} \ (\mu\ne \nu),  \\
G_{(\alpha\beta),(\gamma\gamma)}^{\mu\mu,\mu\mu} & = G_{5L}, \ \ 
G_{(\alpha\beta),(\gamma\gamma)}^{\mu\mu,\nu\nu}  = G_{5T} \ (\mu\ne \nu), 
\end{align}
\label{G123LT}
\end{subequations}
where $\alpha, \beta, \gamma$ and $\delta$ are all different.
Considering various matrix elements of both sides of $Z G = I$ we get

\begin{widetext}
\begin{subequations}
\label{Gcalc-vector}
\begin{multline}
P_L G_{1L} +(m-1) P_T G_{1T} + 2(n-2)(Q_L G_{2L} +(m-1) Q_T G_{2T}) +
\smfrac{1}{2}(n-2)(n-3) (R_L G_{3L} + (m-1) R_T G_{3T}) + \\
2(S_L G_{4L} + (m-1) S_T G_{4T}) +
(n-2)(T_L G_{5L} + (m-1) T_T G_{5T}) = 1
\end{multline}
\begin{multline}
P_L G_{1T} + P_T G_{1L} + (m-2) P_T G_{1T} +
2(n-2)(Q_L G_{2T} + Q_T G_{2L} + (m-2) Q_T G_{2T}) +  \\
\smfrac{1}{2}(n-2)(n-3) (R_L G_{3T} + R_T G_{3L} + (m-2) R_T G_{3T}) +  \\
2(S_L G_{4T} + S_T G_{4L} + (m-2) S_T G_{4T}) + (n-2)(T_L G_{5T} + T_T G_{5L}
+ (m-2) T_T G_{5T}) = 0
\end{multline}
\begin{multline}
Q_L G_{1L} +(m-1) Q_T G_{1T} +
(P_L + (n-2) Q_L + (n-3) R_L) G_{2L} +
(m-1) (P_T + (n-2) Q_T + (n-3) R_T) G_{2T} + \\
((n-3) Q_L + \smfrac{1}{2}(n-3)(n-4) R_L) G_{3L} +
(m-1) ((n-3)Q_T + \smfrac{1}{2}(n-3)(n-4) R_T) G_{3T} + \\
2(S_L G_{4L} + (m-1) S_T G_{4T}) +
(n-2)(T_L G_{5L} + (m-1) T_T G_{5T}) = 0
\end{multline}
\begin{multline}
Q_L G_{1T} + Q_T G_{1L} + (m-2) Q_T G_{1T} +
(P_L + (n-2) Q_L + (n-3) R_L) G_{2T} +
(P_T + (n-2) Q_T + (n-3) R_T) G_{2L} + \\
(m-2) (P_T + (n-2)Q_T + (n-3) R_T) G_{2T} + 
((n-3) Q_L + \smfrac{1}{2}(n-3)(n-4) R_L) G_{3T} + \\
((n-3) Q_T + \smfrac{1}{2}(n-3)(n-4) R_T) G_{3L} +
(m-2) ((n-3)Q_T + \smfrac{1}{2}(n-3)(n-4) R_T) G_{3T} +  \\
2(S_L G_{4T} + S_T G_{4L} + (m-2) S_T G_{4T}) + (n-2)(T_L G_{5T} + T_T G_{5L}
+ (m-2) T_T G_{5T}) = 0
\end{multline}
\begin{multline}
R_L G_{1L} +(m-1) R_T G_{1T} +
(4 Q_L + 2(n-4) R_L) G_{2L} +
(m-1) (4 Q_T + 2(n-4) R_T) G_{2T} + \\
(P_L + 2(n-4) Q_L + \smfrac{1}{2}(n-4)(n-5) R_L) G_{3L} +
(m-1) (P_T + 2(n-4)Q_T + \smfrac{1}{2}(n-4)(n-5) R_T) G_{3T} + \\
2(S_L G_{4L} + (m-1) S_T G_{4T}) +
(n-2)(T_L G_{5L} + (m-1) T_T G_{5T}) = 0
\end{multline}
\begin{multline}
R_L G_{1T} + R_T G_{1L} + (m-2) R_T G_{1T} +
(4 Q_L + 2(n-4) R_L) G_{2T} +
(4 Q_T + 2(n-4) R_T) G_{2L} +
(m-2) (4 Q_T + 2(n-4) R_T) G_{2T} + \\
(P_L + 2(n-4) Q_L + \smfrac{1}{2}(n-4)(n-5) R_L) G_{3T} + \\
(P_T + 2(n-4) Q_T + \smfrac{1}{2}(n-4)(n-5) R_T) G_{3L} +
(m-2) (P_T + 2 (n-4)Q_T + \smfrac{1}{2}(n-4)(n-5) R_T) G_{3T} +  \\
2(S_L G_{4T} + S_T G_{4L} + (m-2) S_T G_{4T}) + (n-2)(T_L G_{5T} + T_T G_{5L}
+ (m-2) T_T G_{5T}) = 0
\end{multline}
\end{subequations}

Forming appropriate linear combinations of Eqs.~\eqref{Gcalc-vector} gives
\begin{multline}
\Bigl([G_{1L} + (m-1) G_{1T}] - 2[G_{2L} + (m-1) G_{2T}] 
+ [G_{3L} + (m-1) G_{3T}] \Bigr) \times \\
\Bigl([P_L + (m-1) P_T] - 2[Q_L + (m-1) Q_T] 
+ [R_L + (m-1) R_T] \Bigr) = 1 .
\label{GLTPQRLT}
\end{multline}
so the ``replicon'' propagator is given by 
\begin{multline}
G_r \equiv 
[G_{1L} + (m-1) G_{1T}] - 2[G_{2L} + (m-1) G_{2T}]
+ [G_{3L} + (m-1) G_{3T}] \\
= \Bigl(  
[P_L + (m-1) P_T] - 2[Q_L + (m-1) Q_T] 
+ [R_L + (m-1) R_T]\Bigr)^{-1} \, .
\label{G-vector}
\end{multline}
\end{widetext}
\section{Averages over spin directions}
\label{sec:averages}
To evaluate the spin glass susceptibility we
need to compute averages over spin directions. Consider, for example,
\begin{equation}
Z = \int d\,\Omega_m  \exp\left[ \mathbf{H}\cdot \mathbf{e}\right],
\label{Zint}
\end{equation}
where the integral is over the surface, $\Omega_m$, of a sphere of unit
radius, $\mathbf{e}$ is a unit vector whose direction is to be integrated over,
and $\mathbf{H}$ is a fixed vector.

Working in polar coordinates, with the polar axis along the direction of the
fixed vector $\mathbf{H}$, the integral in Eq.~\eqref{Zint} can be expressed
entirely in terms of the polar angle $\theta$, since
$\exp\left[ \mathbf{H}\cdot \mathbf{e}\right] = \exp\left[ H \cos\theta \right]$ and
$\int d\,\Omega_m = C_m \int_0^\pi \sin^{m-2}\theta$ for a constant $C_m$. 
To determine $C_m$ we note the
following results\cite{surface,abramowitz:72},
\begin{align}
\Omega_m \equiv \int d\, \Omega_m &= {2 \pi^{m/2} \over \Gamma\left(m\over 2\right)} ,
\label{Omega_m} \\
\int_0^\pi \sin^{m-2}\theta \, d\theta &= \sqrt{\pi}\,\,
{\Gamma\left({m-1\over 2}\right) \over \Gamma\left({m \over 2}\right)} \, ,
\end{align}
where $\Gamma$ is the usual Gamma function,
which gives
\begin{equation}
C_m = {2 \pi^{(m-1)/2} \over \Gamma\left({m-1\over 2}\right)} \, .
\end{equation}
Hence $Z$ can be written as
\begin{equation}
Z = {2 \pi^{(m-1)/2} \over \Gamma\left({m-1\over 2}\right)}  \int_0^\pi 
\exp\left[ H \cos\theta\right] \, \sin^{m-2}\theta\, d \theta \, .
\end{equation}
The integral is given in terms of a modified Bessel
function\cite{abramowitz:72},
and we have
\begin{equation}
Z = (2 \pi)^{m/2} \, {I_{m/2-1}(H) \over H^{m/2-1}} \, .
\end{equation}

Of greater interest are averages of the spins. Consider first
\begin{align}
\langle S_\mu \rangle &= m^{1/2} \langle e_\mu \rangle
, \\
&= m^{1/2}\, {1 \over Z}\,  {\partial Z \over \partial H^\mu} , \\
&= m^{1/2}\, {1 \over Z} \,{H^\mu \over H}\, {\partial Z \over \partial H} .
\end{align}
Using\cite{abramowitz:72}
\begin{equation}
{d \over d x} \left[ I_{m/2-1}(x) \over x^{m/2-1}\right] =  {I_{m/2}(x) \over
x^{m/2-1}} ,
\label{deriv}
\end{equation}
we get
\begin{equation}
\langle S_\mu \rangle =  m^{1/2}\, {H^\mu \over H}\, {I_{m/2}(H) \over
I_{m/2-1}(H)} \, .
\label{Smu}
\end{equation}

We shall also need
\begin{align}
\langle S_\mu S_\nu \rangle  &=  m\, {1 \over Z}
\,{H^\mu \over H}\, {\partial \over \partial H}\, \left( \,{H^\nu \over H}\,
{\partial Z \over \partial H} \right) , \\
&= {m \over I_{m/2-1}(H)}\! \left[ \delta_{\mu\nu} {I_{m/2}(H) \over H} +
{H_\mu H_\nu \over H^2} \, I_{m/2+1}(H) \right] \!,
\label{SmuSnu}
\end{align}
in which we again used Eq.~\eqref{deriv}.

To apply these results, we note that, in the presence of a external random
field, the replica symmetric solution predicts that
$\mathbf{H}$ is given by
\begin{equation}
\mathbf{H} = \beta m^{1/2} \left(J^2 q + h_r^2 \right) \, \mathbf{z} .
\label{H}
\end{equation}
where
each component of the variable $\mathbf{z}$ 
is a Gaussian random variable with zero mean and standard deviation unity.
To see this, compare
Eq.~\eqref{Salpha_av-vector} with Eq.~\eqref{Zint} and note that the spins are of
length $m^{1/2}$ according to Eq.~\eqref{norm}.
Hence each component of $\mathbf{H}$ has zero mean and standard deviation
given by
\begin{equation}
\Delta = \beta m^{1/2} \left(J^2 q + h_r^2\right)^{1/2} \, .
\label{Delta}
\end{equation}

As for the Ising case, we denote averages over $\mathbf{H}$, or equivalently
over $\mathbf{z}$ ($\mathbf{H}$ and $\mathbf{z}$ are related by Eq.~\eqref{H}),
by $[ \cdots ]\z$
and so, for example, in situations which only involve the magnitude of
$\mathbf{H}$,
we have
\begin{align}
[ f(H) ]\z &= 
\int_{-\infty}^\infty\!\! \Bigl(\prod_\mu{d H_\mu
\over (2\pi)^{1/2} \Delta} \Bigr) e^{-\sum_\mu H_\mu^2 / 2
\Delta^2} f(H)\, d H ,
\nonumber \\
&= {\Omega_m \over (2\pi)^{m/2}\Delta^m}  \!\!
\, \int_0^\infty \!\!\!
H^{m-1} \exp\left(-{H^2\over 2 \Delta^2}\right)\!\! \, f(H) \, d H ,
\nonumber \\
&= {2^{1-m/2} \over \Delta^m \Gamma\left({m\over 2 }\right)}
\, \int_0^\infty \!\!\!
H^{m-1}\, \exp\left(-{H^2\over 2 \Delta^2}\right)\!\! \, f(H) \, d H , 
\label{Hav} \\
&= {2^{1-m/2} \over \Gamma\left({m\over 2 }\right)}
\, \int_0^\infty \!\!\!
z^{m-1}\, e^{-z^2/2} \, f(\Delta\, z) \, d z ,
\label{zav}
\end{align}
where we used the result for $\Omega_m$ in Eq.~\eqref{Omega_m}, and
$\Delta$ is given by Eq.~\eqref{Delta}.

Using these results we now calculate the
spin glass order parameter $q$, which is given by
\begin{align}
q &= {1 \over m}\, \left[ \sum_{\mu=1}^m \langle S_\mu \rangle^2 \right]\z \nonumber \\
&= \sum_{\mu=1}^m {(H^\mu)^2 \over H^2}\, \left[\left({I_{m/2}(H) \over
I_{m/2-1}(H)}\right)^2 \right]\z , \nonumber \\
&= \left[\left({I_{m/2}(H) \over
I_{m/2-1}(H)}\right)^2 \right]\z , \nonumber \\
&= {2^{1-m/2} \over \Delta^m \Gamma\left({m\over 2 }\right)}
\, \int_0^\infty d H\,
H^{m-1}\, \exp\left(-{H^2\over 2 \Delta^2}\right)
\left({I_{m/2}(H) \over
I_{m/2-1}(H)}\right)^2 , \nonumber \\
&= {2^{1-m/2} \over \Gamma\left({m\over 2 }\right)}
\, \int_0^\infty d z\,
z^{m-1}\, e^{-z^2/2}\, 
\left({I_{m/2}(\Delta\, z) \over
I_{m/2-1}(\Delta\, z)}\right)^2 , \nonumber \\
\label{q}
\end{align}
where we used Eq.~\eqref{Smu}. Equation \eqref{q}, with $\Delta$ given by
Eq.~\eqref{Delta}, is the self-consistent equation which determines $q$.
As an example, 
for $m=1, I_{m/2}(H) /
I_{m/2-1}(H)= \tanh(H) = \tanh(\Delta\, z)$, and we recover the result for $q$ in
Eq.~\eqref{q_RS}. For general $m$,
expanding the Bessel functions for small
argument\cite{abramowitz:72}, we get
\begin{multline}
q = \Bigl[ {1 \over m^2} H^2 - {2 \over m^3(m+2)} H^4 + \\
{5m + 12 \over m^4(m+2)^2(m+4)} H^6 + O (H^8) \Bigr]\z
\label{q-series}
\end{multline}
If we do the Gaussian integrals, set $h_r = 0$, and solve for $q$, we find
\begin{equation}
q  = t + {1 \over m + 2}\, t^2 + O(t^3)\, , \quad (h_r = 0)\, , 
\label{q=t}
\end{equation}
where $t$, the reduced temperature, is given by $t  = (T_c - T)/T_c$,
and the zero field transition temperature is $T_c = J$,
see Eq.~\eqref{Tc}.

Our main goal is to compute the eigenvalue $\lambda_{3S}$ since this determines
the spin glass susceptibility, the divergence of which indicates the
location of the AT line. 
From Eqs.~\eqref{l3vecS}, \eqref{A}, \eqref{PLPT}, \eqref{Z-vector} and 
\eqref{Z},
we find the fairly simple expression
\begin{multline}
\lambda_{3S} = 1 - (\beta J)^2 \, {1\over m}\, \sum_{\mu,\nu} \\
\left[
\langle S_\mu S_\nu\rangle^2 - 2 \langle S_\mu S_\nu\rangle 
\langle S_\mu \rangle  \langle S_\nu \rangle +
\langle S_\mu \rangle^2  \langle S_\nu \rangle^2
\right]\z ,
\label{lambda3S}
\end{multline}
which is instructive to write in the following form
\begin{equation}
\lambda_{3S} = 1 - (\beta J)^2 \, \chi_{SG}^0 ,
\label{lambda3S0}
\end{equation}
where $\chi_{SG}^0$ is a single-site spin glass susceptibility,
\begin{equation}
\chi_{SG}^0 =  {1\over m}\, \sum_{\mu,\nu} 
\left[ \Bigl(
\langle S_\mu S_\nu\rangle - 
\langle S_\mu \rangle  \langle S_\nu \rangle \Bigr)^2
\right]\z . 
\label{chiSG_0}
\end{equation}

Evaluating the spin averages in Eq.~(\ref{chiSG_0}) using
Eqs.~\eqref{Smu} and \eqref{SmuSnu} gives
\begin{widetext}
\begin{multline}
\chi_{SG}^0 = m 
\Bigl[ {1 \over I^2_{m/2-1}(H)}
\left\{ {m \over H^2} I^2_{m/2}(H) + {2 \over H} I_{m/2}(H) I_{m/2+1}(H) +
I^2_{m/2+1}(H)  \right\} -
\\
{2 \over I^3_{m/2-1}(H)} \left\{
{1 \over H} I^3_{m/2}(H) + I^2_{m/2}(H) I_{m/2+1}(H) \right\} + 
\left\{ {I_{m/2}(H) \over I_{m/2-1}(H)}\right\}^4 
\Bigr]\z .
\label{l3s_m}
\end{multline}
\end{widetext}
We recall that the average over $H$
is evaluated according to Eq.~\eqref{Hav}. 
For the Ising case, $m = 1$, 
Eq.~\eqref{l3s_m} simplifies to
\begin{equation}
\chi_{SG}^0 = 
\left[ 1 - 2 \tanh^2 H + \tanh^4 H \right]\z,
\label{l3_Ising}
\end{equation}
in agreement with Eq.~\eqref{chisg0_Ising}.
For the Heisenberg case, $m=3$, Eq.~\eqref{l3s_m} becomes
\begin{equation}
\chi_{SG}^0 = 3
\left[{3 + 2 H^2 - 4 H \coth(H) \over H^4}
+ {1 \over \sinh^4(H)}
\right]\z \!\!\!,
\label{l3s_Heis}
\end{equation}
which, together with Eqs.~\eqref{Hav} and \eqref{lambda3S0}, gives $\lambda_{3S}$.
Equations \eqref{l3s_m} and \eqref{l3s_Heis} appear to be a new results.
Expanding the Bessel functions for small $H$ gives
\begin{equation}
\chi_{SG}^0 = \left[1 - {2\over m^2} H^2 + {5m + 16 \over m^3(m+2)^2}H^4 +
O(H^6) \right] \, .
\label{chisg-series}
\end{equation}

Let us evaluate $q$ and $\lambda_{3S}$ near $ T = T_c \ (= J), $
the zero field
transition temperature, and for small
$h_r$.
Using Eqs.~\eqref{q-series} and \eqref{chisg-series},
and doing the Gaussian integrals,
we find
\begin{align}
q &= \widetilde{\Delta}^2 - 2  \widetilde{\Delta}^4 + {5m + 12 \over m + 2} \,
\widetilde{\Delta}^6 + \cdots \, ,\label{qseries}\\
\lambda_{3S} &= 1 - (\beta J)^2 \left[ 1 - 2 \widetilde{\Delta}^2 +
{5m + 16 \over m + 2} \widetilde{\Delta}^4 + \cdots \, \right] \, ,
\label{l3series}
\end{align}
where
\begin{equation}
\widetilde{\Delta}^2 \equiv {\Delta^2 \over m} = \beta^2(J^2 q + h_r^2) \, .
\end{equation}
Combining Eqs.~\eqref{l3series} and \eqref{qseries}, and assuming 
\begin{equation}
h_r \ll 
t \equiv (T_c - T)/T_c \ll 1\, ,
\label{limits}
\end{equation}
which will be valid at and below the AT line near $T_c$, we
get
\begin{equation}
\lambda_{3S} = \left({h_r \over J}\right)^2 \, {1 \over q}
- {4 \over m + 2}\, q^2 \, .
\end{equation}
In the limits of Eq.~\eqref{limits},
we have
$q = t + O(t^2)$, see Eqs.~\eqref{q=t} and \eqref{qAT}, 
and so
\begin{equation}
\lambda_{3S} = \left({h_r \over J}\right)^2 \, {1 \over t}
- {4 \over m + 2}\, t^2 \, , \quad (h_r \ll t) \, ,
\label{unstable}
\end{equation}
which changes sign for
\begin{equation}
\left( {h_r \over J}\right) ^2 = {4 \over m + 2}\, t^3 .
\label{HrAT}
\end{equation}
Equation \eqref{HrAT}
gives the location of the AT line for an $m$-component spin glass near
the zero field transition.
The replica symmetric solution is unstable at lower
temperatures and fields since $\lambda_{3S} < 0$ in that region
according to Eq.~\eqref{unstable}.
Note that Eq,~\eqref{HrAT} correctly gives the AT result
that $h_r^2 =  (4/3)\,t^3$ for $m = 1$ (This is
a valid comparison even though AT used a uniform field
since, to lowest order in $t$, the location of
the AT line in the Ising case is the same\cite{bray:82b} for random
and
uniform fields.)
On the AT line we find that the spin glass order parameter is
given by
\begin{equation}
q = t + {3 \over m+ 2}\, t^2 + O(t^3)\, , \quad \text{(on AT line)} \, .
\label{qAT}
\end{equation}

In the opposite limit, $T \to 0$, we find, using properties of the Bessel
functions, that
\begin{equation}
{h_r(T = 0) \over J} = {1 \over \sqrt{m-2}} \quad (m > 2) \, ,
\label{lowTm}
\end{equation}
while $h_r(T \to 0)$ diverges for $m \le 2$. For the Ising case,
we get
\begin{equation}
{h_r(T\to0) \over J}  = \sqrt{8 \over 9 \pi} \,\, 
{J \over T} \quad (m=1) \, ,
\label{lowTising}
\end{equation}
in agreement with Bray\cite{bray:82b}.

\bibliography{refs,comments}

\begin{thebibliography}{36}
\bibitem[{mis()}]{missing}
\bibinfo{note}{The $(\beta J)^{-2}$ term in Eq.~\eqref{chisg_qab}, which comes
  from the delta function in Eq.~\eqref{Sabcd'}, was inadvertently omitted in
  Eq.~(4.47) of Ref.~\onlinecite{binder:86}, though it was present in
  Eq.~(4.45) and (implicitly) in Eq.~(4.52) of that reference.}
\end{thebibliography}

\end{document}